\documentclass[journal]{IEEEtran}

\usepackage{array,multirow,graphicx}
\usepackage{float}

\usepackage{algorithmicx}
\usepackage[ruled]{algorithm}
\usepackage{algpseudocode}
\usepackage{cite}

\usepackage[english]{babel}
\usepackage[latin1]{inputenc}

\usepackage{epsfig}
\usepackage{color}
\usepackage{subcaption}

\usepackage{times}
\usepackage{url}
\usepackage{amsthm}
\usepackage{amsmath}
\usepackage{amssymb}
\usepackage{dsfont}
\usepackage[disable]{todonotes} 

\graphicspath{{../../../../Simulaciones/cens5d/Figs/}}                   
\theoremstyle{definition} 
\newtheorem{corollary}{Corollary} 
\newtheorem{lemma}{Lemma} 
\newtheorem{theorem}{Theorem}

\newcommand{\prob}{\mathbb{P}}
\newcommand{\Ex}{\mathbb{E}}
\newcommand{\Var}{\mathbb{V}\mathrm{ar}}

\newcommand{\R}{\mathbb{R}}

\newcommand{\Hip}{\mathcal{H}}
\newcommand{\N}{\mathcal{N}}

\newcommand{\diag}{\text{diag}}
\newcommand{\tr}{\text{tr}}
\newcommand{\ve}[1]{\boldsymbol{#1}}
\newcommand{\mat}[1]{\mathbf{#1}}

\newcounter{mycomment}

\begin{document}

\title{On fully-distributed composite tests with general parametric data distributions in sensor networks}
\author{{Juan Augusto Maya~\IEEEmembership{Member,~IEEE,} and Leonardo Rey Vega,~\IEEEmembership{Member,~IEEE.}}
	\thanks{The two authors are with Universidad de Buenos Aires, Buenos Aires, Argentina, and CSC-CONICET, Buenos Aires, Argentina. 
		Emails:  \{jmaya, lrey\}@fi.uba.ar. This work was partially supported by the projects UBACyT 20020170200283BA and PICT-2017-4533. } %
} 
\maketitle
\begin{abstract}
We consider a distributed detection problem where measurements at each sensor follow a general parametric distribution. The network does not have a central processing unit or fusion center (FC). Thus, each node takes some measurements, does some processing, exchanges messages with its neighbors and finally makes a decision (typically the same for all nodes) about the phenomenon of interest. The problem can be formulated as a composite hypothesis test with unknown parameters where, in general, a uniformly most powerful test does not exist. This leads naturally to the use of the Generalized Likelihood Ratio (GLR) test. As the measurements follow a general parametric distribution (which could model spatial dependence of the data), the implementation of fully-distributed detection procedures could be demanding in network resources. For this reason, we study the use of a simpler test (referred as L-MP) which uses the product of the marginals of the measurements taken at each node, where the unknown parameters are easily estimated with only local measurements. Although this simple proposal still requires network-wide cooperation between nodes, the number of communications is significantly reduced with respect to the GLR test, making it a suitable choice in severely resource-constrained sensor networks. This simpler test does not exploit the full parametric model of data, so, it becomes important to analyze its statistical properties and its potential performance loss. This is done through the analysis of the L-MP asymptotic distribution. Interestingly, despite the fact that the L-MP is simpler and more efficient to implement than the GLR test, we obtain some conditions under which the L-MP has superior asymptotic performance to the GLR test. Finally, we present numerical results for a fully-distributed spectrum sensing application for cognitive radios, showing the gains of the L-MP in terms of performance, and saving of resources in comparison with some other well-known approaches for this application.
\end{abstract}

\section{Introduction}
In the near past, Wireless Sensor Networks (WSN) have received considerable attention from the research and industrial community because of their remote monitoring and control capabilities \cite{shaikh2016energy, rashid2016applications,li2020distributed}. More recently, they have become an essential part of the emerging technology of Internet of Things (IoT) \cite{gupta2020collaborative,gubbi2013internet,al2015internet}. Among the different tasks to be done by WSNs, distributed detection is an active research topic \cite{chepuri2016sparse,ciuonzo2017distributed, aldalahmeh2019fusion}.

In a fully-distributed detection problem, geographically distributed sensors collect measurements from the phenomenon of interest, make some processing, exchange information with their neighbors and, finally, execute some consensus or diffusion algorithm to achieve a common final decision. This is different to the more studied case in which there is a FC, which receives the sensor measurements and makes a centralized processing and decision. Although this solution could present some advantages, such as avoiding restrictions to build the desired statistic, it leads to more infrastructure costs and also presents some important vulnerabilities, given the single point of failure introduced by the FC. On the other hand, fully-distributed algorithms are more robust against node failures, and the communications between nodes are done locally, over typically short distances, saving energy and also bandwidth by employing spatial reuse of the frequency bands.

\subsection{Previous results in the literature}
Many works have considered distributed detection architectures but most of them assume the presence of a FC \cite{ciuonzo2014decision, hamed2012reliable, kar2011distributed, li2018fully, Ciuonzo_Rossi_Willett_2017,ghasemi2007asymptotic}. On the other hand, there exists considerable interest in fully-distributed detection architectures without the presence of a FC \cite{sayed2014adaptation,al2018node,sayed2013diffusion}. However, most of these works assume that network-wide measurements have some specific statistical properties. For example, many works consider a binary hypothesis testing problem where the sensors measure noise only under the null hypothesis, and noise plus a known or unknown deterministic signal under the alternative hypothesis, with the noise being independent across the sensors. Thus, the observations at each node are independent conditioned on each hypothesis \cite{Sahu_Kar_2017, cattivelli2011distributed}. Although this is a valid model in several situations, in many applications of interest, the measurements do not verify those conditions. For example, this is the case of spatially dependent measurements \cite{drakopoulos1991optimum}, which appears in WSNs sensing physical variables like temperature and humidity \cite{villas2014spatial}, or acoustic signals \cite{Wang_Reiss_Cavallaro_2016,chen2019layered}. Also, the signals measured under either hypothesis could be more effectively modeled as pure random signals without deterministic components as assumed in several works. For example, this could be the case of devices transmitting communication signals in the context of spectrum sensing for cognitive radio \cite{lunden2015spectrum,Liang_Zeng_Peh_Hoang_2008}. 

On the other hand, some works have considered dependent observations under one of the hypothesis using Gaussian Markov Random Fields \cite{Tong_2007, anandkumar2009scalable} to design a Neyman-Pearson detector with a FC, and analyzing the effect of dependent measurement in the asymptotic regime \cite{Lin_Chen_2015}. More recently, the copula theory \cite{nelsen2007introduction} has been proposed to consider statistically dependent observations, under either hypothesis, recollected by the network. Although this theory allows to deal with general models, the obtained detectors are typically very complex to evaluate and involve multidimensional integrals \cite{he2015fusing,iyengar20178}. That increases the detector complexity and makes it difficult to obtain a fully-distributed detection algorithm, and a central processing unit is typically required. Moreover, the analytical computation of the performance, even in the asymptotic regime, is very challenging. 

Other algorithms, originally developed for multi-antenna systems, are based on the eigenvalues of the sample covariance matrix of the observations, and their use is very extended in the literature for detection problems with correlated measurements \cite{zhang_multi-antenna_2010, taherpour_multiple_2010, penna_decentralized_2015}. However, the typical computation cost of them in a fully distributed manner is very high in terms of network communications resources \cite{penna_decentralized_2015}.
Thus, the design of distributed detection algorithms with arbitrary measurements in a fully-decentralized scenario deserves more investigation \cite{javadi2016detection}. 

\subsection{Contributions and results}
In this work we deal with a composite hypothesis testing problem where sensors take measurements distributed under a general parametric distribution for both hypotheses. This model is sufficiently general and allows to capture the main characteristics of the data in different scenarios. As it is assumed that there is no FC in the network, each node should reach, via some consensus mechanism, a decision about the true state of nature (e.g., if a source signal is present or not). Besides this, it is also assumed that some parameters of the data distribution are unknown, which precludes the use of the optimal Neyman-Pearson detector. 
Typically unknown parameters are the channel gain between the source and the receivers, the transmit power of the source, and the spatial correlation between measurements of different sensors. 
As in the case of multidimensional measurements there are no guarantees of the existence of a uniformly most powerful test, the approach of the generalized likelihood ratio (GLR) test, where the unknown parameters are estimated using the maximum likelihood estimator (MLE), is a potential approach to the problem. However, because of the fully-decentralized nature of the network, the implementation of this approach for general parametric measurement distributions is not straightforward. This is because, even in the case in which the parameters are known, the resulting optimal Neyman-Pearson test could require numerous communication exchanges between the nodes for building the statistic in a fully-distributed fashion. The message exchanges require communication resources (such as transmit power and bandwidth) which are typically scarce (at least in WSNs). In this sense they should be limited as much as possible with minimal impact on the performance of the detection task. Moreover, in the case of distributions with unknown parameters, we should also consider the formation of the MLE estimator. Although it is possible to implement a MLE in a decentralized scenario, it generally also requires a high cost of communication resources. This is even more critical if there is not a closed-form expression for the MLE estimator. In that situation, the nodes should cooperatively implement a numerical procedure to find the estimator, leading to even more exchanges. To alleviate this cost, we study a decision statistic (referred as L-MP) that uses the product of the marginal PDFs. This is relevant because the estimation procedure of the unknown parameters of the probability distributions can be done locally at each sensor node. Besides that, the resulting statistic has an attractive factorization structure that facilitates its computation in a distributed scenario, resulting in low implementation costs in terms of network resources.

As the L-MP does not use the full parametric model in order to implement the decision statistic, some loss of performance could be expected (e.g. because of not exploiting possible spatial correlation of the measurements). Then, we need to characterize the performance and evaluate possible penalties introduced by the proposed strategy. To this end, we derive the theoretical asymptotic distribution of the L-MP which allows us to compute the error probabilities (type I or II) of the test. Although these results are strictly valid in the asymptotic scenario, when the number of measurements tends to infinity, we show with numerical examples that they offer a good matching also in the finite length regime. Surprisingly, in some scenarios of practical interest, the L-MP performs better than other approaches, including the GLR test, which explicitly takes into account the full parametric model of the measurements. Some conditions are also presented that allow us to explain the main reasons for that behavior. Additionally, as one of the main concerns in distributed detection applications is the consumption of network resources, we also perform an analysis of the communication costs of the L-MP scheme showing its significant savings with respect to other solutions.

The paper is organized as follows. In the next section we introduce the general model of the problem. In Sec. III we present the studied fully-distributed statistic and in Sec IV, we derive its asymptotic performance and discuss some important results regarding the performance of the fully-distributed test and the GLR one. An analysis of the communications resources for the implementation of different distributed detection schemes is also presented. In Sec V, we evaluate the studied fully-distributed statistic and other competing solutions in the context of spectrum sensing for cognitive radio networks. We finally draw the main conclusion in Sec. VI. We leave the proof of the asymptotic results to the appendix.

\subsubsection*{Notation}
Vectors and matrices are written in bold letters. $\mat I_P$ is the identity matrix of size $P\times P$. The gradient of a scalar function $p$ with respect to (wrt) the vector $\ve x\in \R^{M_1}$ is noted as $\frac{\partial p}{\partial\ve x}$ and assumed to be a column vector. If $\ve y\in \R^{M_2}$, $\frac{\partial^2 p}{\partial\ve x\partial \ve y}$ is a $M_1\times M_2$ matrix with its $(i,j)$-th component being $\frac{\partial^2 p}{\partial x_i\partial y_j}$, $i\in [1:M_1]$, and $j\in[1:M_2]$. $\Ex_{\ve \phi}(\cdot)$ denotes the expectation wrt the PDF $p(\cdot,\ve \phi)$ with parameter $\ve\phi$.     
A Gaussian vector $\ve n $ with mean $\ve\mu$ and covariance matrix $\ve{\Sigma}$ is notated $\ve n\sim \N(\ve \mu,\ve\Sigma)$.  $\text{diag}(\mat A)$ and $\text{diag}(\ve a)$ are diagonal matrices built with the diagonal elements of the square matrix $\mat A$, or the elements of the vector $\ve a$, respectively. $\bar{\text{diag}}(\mat A)$ is a vector built with the diagonal elements of $\mat A$, and $\text{vech}(\mat A)$ is an operator that concatenates all but the supra-diagonal elements of $\mat A$. For a given statistic $T$ and a predefined threshold $\tau$, the miss-detection and false alarm probabilities will be denoted, respectively, $P_\text{md}=\prob(T<\tau|\Hip_1)$, and $P_\text{fa}=\prob(T>\tau|\Hip_0)$.
\section{Model}
\label{sec:model}
We consider a composite binary hypothesis testing problem in a WSN with $N$ nodes. Assume that each sensor takes $L$ observations, which are statistically independent and identically distributed (iid) in time, a common assumption considered in the literature of distributed detection (e.g., see  \cite{sayed2013diffusion, li2018fully} and references therein), but possibly dependent across the sensors. Let ${\ve z}_l\equiv[{z}_{1}(l),\dots,{z}_{N}(l)]^T\in\R^N$ be the observations taken by all nodes at the $l$-th slot time, $l\in[1:L]$, and $\ve z \equiv \{\ve z_1,\dots,\ve z_L\}$ which contains all the network measurements. 
We assume that the hypothesis testing problem can be expressed as a parameter test \cite{Kay_SSP}. 
We let the joint PDF of $\ve z_l$ with the global vector parameter $\ve\theta\in \R^M$ be $p(\ve z_l;\ve \theta)$. The true vector parameter under the hypothesis $\Hip_i$ is $\ve\theta^i$, $i=0,1$. 
The test is
\begin{equation}
\Hip_0:{\ve z}_l \overset{iid}{\sim} p({\ve z}_l; \ve \theta^0 ),  
\Hip_1: {\ve z}_l \overset{iid}{\sim} p({\ve z}_l; \ve \theta^1 ), \ l\in[1:L], 
\label{eq:htpeq}
\end{equation}
equivalently expressed as a parameter test as
\begin{equation}
\Hip_0:\ve\theta=\ve\theta^0, \Hip_1:\ve\theta\neq\ve\theta^0,
\label{eq:param-test-g}
\end{equation}
where we assume that $\ve\theta^0$ is known, and $\ve\theta^1\neq\ve\theta^0$ is unknown. We define the local vector parameter $\ve \theta_k^\text{loc}\in \R^{M_k}$, $k\in[1:N]$, as the parameter that completely describes de marginal PDF of the $k$-th node, i.e.,
$\int\cdots\int p(\ve z_l;\ve \theta)dz_1(l)\dots dz_{k-1}(l)dz_{k+1}(l)\dots dz_{N}(l) = p_k(z_k(l);\ve\theta_{k}^\text{loc}).$ We let  $\ve{\theta}^\text{loc}\equiv\{\ve{\theta}_1^\text{loc},\dots,\ve{\theta}_N^\text{loc}\}$, $\ve{\theta}^\text{loc}\in\R^{P}$, $P=\sum_{k=1}^N M_k$. 
$\ve{\theta}^\text{loc}$ is the set of parameters that are \emph{observable} at individual nodes and can be estimated locally without knowledge of the samples taken at other nodes. We observe that $\ve{\theta}^\text{loc}$ is a function of $\ve{\theta}$, although usually two situations are common. In the first one $\ve{\theta}^\text{loc}$ is a subset of $\ve{\theta}$. For example, if $N=2$ and $\ve\theta=\{\theta_1,\theta_2,\theta_3\}$ with $M=3$, we could have $\theta_1^\text{loc}=\theta_1$, and $\theta_2^\text{loc}=\theta_2$, where $\theta_3$ is a global parameter non-observable locally at any node. This parameter could be for example the spatial correlation coefficient between the measurements of two different nodes. In the second one, $\ve{\theta}^\text{loc}$ has some global parameters repeated as local parameters at each node. For example, we could have $\ve\theta_1^\text{loc}=\{\theta_1,\theta_3\}$ and  $\ve\theta_2^\text{loc}=\{\theta_2,\theta_3\}$, where $\theta_3$ is a global parameter that is observable at both nodes. For example, it could be the variance of the sensed noise when the nodes are similar electronic devices.	
\begin{figure}[b]
	\centering
	\includegraphics[width=1\linewidth]{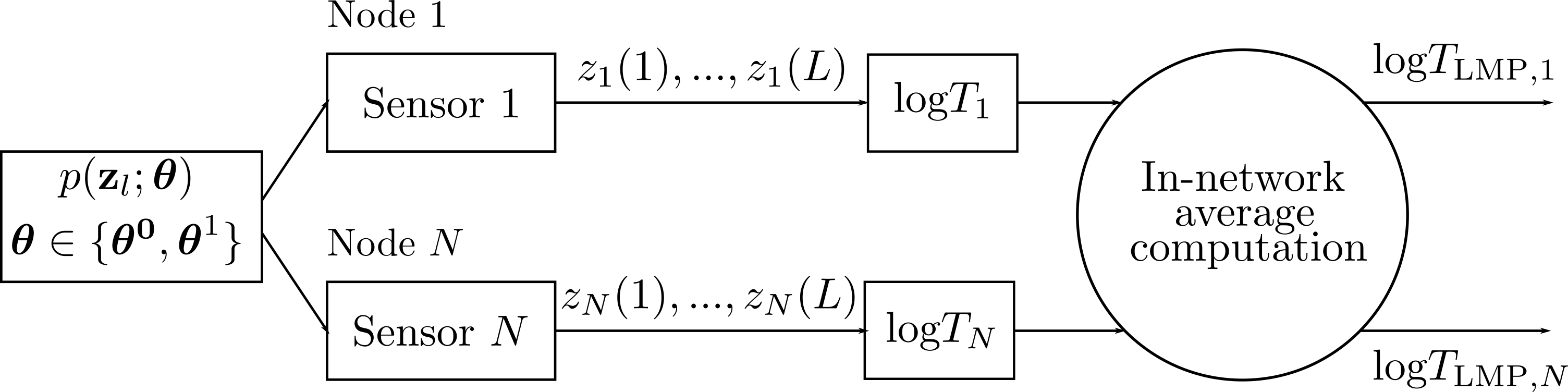}
	\caption{Schematic representation of the WSN for implementing a fully-distributed algorithm for detecting $\Hip_0$ or $\Hip_1.$}
	\label{fig:fdd}
\end{figure}

In Fig. \ref{fig:fdd} we show the system model considered for deciding about $\Hip_0$ or $\Hip_1$. In the next section, we define the fully-distributed statistic and how it is computed through a simple in-network averaging procedure using a consensus algorithm.

\section{Statistics under study} 
\label{sec:D-GLRT}
\subsection{A brief review of the GLR test}
To perform the test (\ref{eq:htpeq}) we consider the GLR test approach, frequently used in the literature and whose asymptotic performance can be computed analytically \cite{Kay_SSP,Levy_Det}.
The classical GLR statistic is 
$T_\text{G}(\ve z)\equiv\frac{p(\ve z;\ve \theta = \hat{\ve \theta}_\text{G-MLE})}{p(\ve z; \ve \theta = \ve\theta^0)}$, where $\hat{\ve \theta}_\text{G-MLE}$ is the (global) MLE\footnote{We call it \emph{global} MLE to differentiate it from the local one defined next.} of $\ve \theta_1$.  
The asymptotic distribution of the global GLR statistic $T_\text{G}$ under the so called \emph{weak signal} condition (i.e., there exists a constant $c$ such that $\|\ve\theta^1-\ve\theta^0\|\leq c/\sqrt{L}$) is well known \cite{Kay_SSP}: 
\begin{equation}
2\log T_\text{G}(\ve z)\overset{a}{\sim}\left\{\begin{array}{ll}
\chi^2_M & \text{under } \Hip_0\\
\chi'^2_M(\lambda_g) & \text{under } \Hip_1,
\end{array}\right.
\label{eq:gglr}
\end{equation}
where the symbol $\overset{a}{\sim}$ means ``asymptotically distributed as when $L$ tends to infinity", 
$\chi^2_M$ is the chi-square distribution with $M$ degrees of freedom and $\chi'^2_M(\lambda_g)$ is the non-central chi-square distribution with $M$ degrees of freedom and non-centrality parameter 
$\lambda_g = L(\ve\theta^1-\ve\theta^0)^T \ve i(\ve \theta^0)(\ve\theta^1-\ve\theta^0)$,
where $\ve i(\ve \theta^0)\!$ is the Fisher information matrix at $\ve \theta^0$ \cite{Kay_SSP_ET}. 

\subsection{Local MLE estimate}
In many cases the global MLE is difficult to compute in a distributed scenario. The reason for this is two-fold. In the first place, depending on the structure of $p(\ve z_l;\ve \theta)$, the maximization problem can be very hard from a computational point of view, even in a centralized scenario. On the other hand, computing the global MLE would require in general a large number of sensor communications (of the sensor measurements or some statistics of them) imposing a serious practical constraint in terms of energy, bandwidth and/or delay.
Looking for a simpler approach to compute the MLE we consider a \emph{local} MLE in the sensor node $k$ which only use the locally sensed values $\left\{z_k(l)\right\}_{l=1}^L$, whose distribution under $\Hip_1$ is $\prod_{l=1}^L p_k(z_k(l),\ve\theta_{k}^\text{loc,1})$.
Thus, the $k$-th local parameter estimate $\hat{\ve \theta}_k^\text{loc}$ of $\ve\theta_{k}^\text{loc,1}$ is defined as, $k\in[1:N]$, 
\begin{equation}\textstyle
\hat{\ve\theta}_{k}^\text{loc}\hspace{-5pt}\equiv \arg\max_{\ve\theta_k^\text{loc}}\hspace{-3pt} \frac{1}{L}\sum_{l=1}^{L}\log p_k(z_k(l);\ve\theta_k^\text{loc}).
\label{eq:lmleDef}
\end{equation} 
It is important to note that $\hat {\ve \theta}_k^\text{loc}$ is still an asymptotically consistent estimator of the true parameter $\ve{\theta}_{k}^\text{loc,1}$ given that it is a maximum likelihood estimator. However, the concatenation of all local vector parameters $\ve\theta^\text{loc}$ could lose the property of asymptotic \emph{efficiency} of the global MLE \cite{Kay_SSP_ET} for the same set of parameters, given that $\{\hat {\ve \theta}_k^\text{loc}\}_{k=1}^N$ are estimated through the corresponding marginal distributions instead of the joint probability. Clearly, this estimation efficiency loss could impact negatively in the detection performance of a test that uses the local estimator. Thus, a theoretical performance characterization is important, among others things, to evaluate this effect. We will cover this in the next section.

\subsection{A fully-distributed statistic}
Maximization of the life-time of WSNs is typically a very important goal, therefore, it is critical to minimize the message exchanges between the nodes. This is because the use of the wireless medium is typically expensive in terms of energy and bandwidth. A fully-distributed information processing task could require a large number of exchanges between the nodes. In this sense, it is important to explore strategies that provide communication resource savings and, simultaneously, present reasonable levels of performance for the main task they are intended to solve. Therefore, we propose to build a statistic using the marginal distributions instead of the joint distribution of the data. Specifically, we define the following statistic:
\begin{align}	
\textstyle \log T_\text{L-MP}(\ve z) &\equiv \log\left( \frac{p_\text{MP}(\ve z_l; \ve\theta_{k}^{\text{loc},1} = \hat{\ve\theta}_k^\text{loc} )}{p_\text{MP}(\ve z_l; \ve\theta_k^\text{loc,0} )}\right),\nonumber\\
&= \sum_{k=1}^N \log\frac{ p_k(\{z_k(l)\}_{l=1}^L; \ve\theta_{k}^\text{loc} = \hat{\ve\theta}_k^\text{loc} )}{p_k(\{z_k(l)\}_{l=1}^L; \ve\theta_{k}^\text{loc} = \ve\theta_{k}^\text{loc,0})}
\label{eq:TLMP}
\end{align}
with $p_\text{MP}(\ve z_l; \ve \theta^{\text{loc}}) \equiv\Pi_{k=1}^N p_k(z_k(l); \ve\theta_{k}^{\text{loc}})$ and 
where the subscript refers to the statistic that uses the local estimation of the parameters (L) and the joint PDF is replaced by the product of the marginal PDFs (MP) under each hypothesis. 
We will refer to the following (local) parameter test when we use (\ref{eq:TLMP}) for deciding for one of both hypotheses of the test:     
\begin{equation}
\Hip_0:\ve\theta^\text{loc}=\ve\theta^\text{loc,0}, \Hip_1:\ve\theta^\text{loc}\neq\ve\theta^\text{loc,0}.
\label{eq:param-test-lmp}
\end{equation}
The structure of $T_\text{L-MP}$ opens opportunities to save valuable resources in a WSN such as energy and bandwidth for communicating the quantities required to run the detection algorithms as will be shown also in Sec. IV. Considering $\log T_\text{L-MP}(\ve z)$ we see that each sensor is able to compute its corresponding term in the sum and then, share this quantity to the rest of the sensors to obtain $\log T_\text{L-MP}$ via a simple consensus algorithm. 

\subsection{Spatial averaging}
\label{sec:spatialavg}
The statistic $\log T_\text{L-MP}$ in (\ref{eq:TLMP}) requires the computation of a \emph{spatial} sum $\sum_{k=1}^N (\cdot)$ over all the sensors in the network. Next $\bar{a}\equiv\sum_{k=1}^N a_k$ will represent that sum.
Each sensor node generates locally a scalar value $a_k\in \R$, $k\in\mathcal{N}\equiv[1:N]$ and it is desired to compute the average $\tilde{a} = \frac{1}{N}\sum_{k=1}^N a_k$ (or the sum $\bar{a} = N\tilde{a}$) at each node in a distributed manner and with minimal resources allocated to the exchanges between the nodes. The spatial averages can be computed via a consensus procedure such as in \cite{xiao2004fast, sayed2013diffusion}. 
Consider a network (modeled as a connected graph) $\mathcal{G}=(\mathcal{N},\mathcal{E})$ consisting of a set of nodes $\mathcal{N}$ and a set of edges $\mathcal{E}$, where each edge $\{i,j\}\in\mathcal{E}$ is an unordered pair of distinct nodes. The set of neighbors of node $i$ is denoted by $\mathcal{N}_i = \{j\in\mathcal{N}|\{i,j\}\in \mathcal{E}\}$. 
The average value $\tilde{a}$ can be computed iteratively as,  $t\in\mathbb{N}$:
\begin{equation}\textstyle
a_k(t) = W_{kk} a_k(t-1) + \sum_{j\in\mathcal{N}_k} W_{kj} a_j(t-1),\ \ k\in\N,
\label{eq:spatialAve}
\end{equation}
where $a_k(t)$ is the average after $t$ iterations (or message exchanges between the nodes), $a_k(0)=a_k$ is the initial value and $W_{kj}$ is the weight on $a_j(t-1)$ at the node $k$. 
Considering local transmissions only, i.e., each node broadcasts its local value\footnote{In practice, a quantization scheme is needed but for simplicity it is not considered here.} at iteration $t$ only to the nodes in its neighborhood, we have that for each $k\in\N$, $W_{kj}=0$ for $j\notin\N_k$ and $j\neq k$. 

Among all the existing possibilities for selecting the weights, we will consider a simple but effective algorithm called \emph{local-degree weights} distributed averaging algorithm \cite{xiao2004fast}. Its convergence to the required average is guaranteed given that the graph is not bipartite. 
The weights are assumed to be symmetric with value $W_{kj}=W_{jk}=1/\max(d_k,d_j)$, where $d_k$ is the degree of node $k$, i.e., the number of neighbors of the node $k$. 
Algorithm \ref{fd} summarizes the steps required to compute the statistic $T_\text{L-MP}$. 
Several stopping criteria can be considered for the iterative computation of the spatial average (\ref{eq:spatialAve}). Here we consider a fixed number of exchanges $N_{it}$. 
\begin{algorithm}[t]
\caption{Distributed implementation of $T_{\text{L-MP}}$}\label{fd}
\begin{algorithmic}[1]
\For {$k=1,\dots,N$}{ (\it simultaneously at each sensor)}
\State Compute the local estimate $\hat{\ve \theta}_{k}^\text{loc}$ using eq. (\ref{eq:lmleDef}).
\State Compute $\log T_k\equiv\log\frac{p_k(\{z_k(l)\}_{l=1}^L; \theta_{k} = \hat{\theta}_{k}^\text{loc})}{p_k(\{z_k(l)\}_{l=1}^L; \theta_{k} = \theta_{k}^{\text{loc},0})}$.
\State $\log T_{\text{L-MP},k}=$ \Call{SpatialSum}{$\{\log T_j\}_{j\in\N_k\cup\{k\} }$}
\If {$\log T_{\text{L-MP},k}<\gamma$} {Sensor $k$ decides $\Hip_0$,} 
\Else { Sensor $k$ decides $\Hip_1$.} 
\EndIf
\Comment{$\gamma$ is the predefined threshold of the test.}  
\EndFor
\Function{SpatialSum}{$\{a_k\}_{k\in\N_k\cup \{k\}}$}
\Comment{Compute $\bar{a}_k$.}
\State $t = 0$, $a_k(0)=a_k$\Comment{Initial condition for $t=0$.}
\While{$t<N_{it}$}
\State $t = t+1$
\State Compute the spatial average $a_k(t)$ using (\ref{eq:spatialAve}).
\EndWhile
\State \textbf{return} $N a_k(t)$ \Comment{Return the sum $\bar{a}_k$}
\EndFunction
\end{algorithmic}
\end{algorithm}

\section{Asymptotic performance analysis}
In this section we present the main results of the paper. First, in Lemma 1, we obtain the asymptotic distribution of the L-MP test. This characterization is important because it allows to theoretically compute the performance of the proposed test in a general scenario, and also to compare it against well-established techniques such as the GLR test. We also find some conditions under which it outperforms the GLR test. This should not be surprising given that the widely used GLR statistic, in general, has not optimality guarantees for composite hypothesis testing problems \cite{Kay_SSP,Levy_Det}, despite the fact that it uses the full parametric structure of $p(\ve z; \ve \theta)$. Then, in Sec \ref{sec:gaussian}, we evaluate the results for the Gaussian case, and finally, in Sec. \ref{sec:resources} we analyze the communication resources needed for computing the L-MP statistic and compare it against other fully-distributed statistics.
\subsection{Theoretical results for general PDFs}
The asymptotic PDF of $T_\text{L-MP}$ is presented next.
The proof is based on classical tools used in the GLR theory, and it is relegated to Appendix A.
\begin{lemma}
	\label{lem:localparam}
	Assume 
		i) the first and second-order derivatives of the log-likelihood function are well defined and continuous functions.
		ii) $\Ex[\partial\log p_k(z_k(l);\ve\theta_k^\text{loc})/\partial\ve\theta_k^\text{loc}]\!=\!\ve 0$, $\forall l$, $k\!\in[1:\!N]$.
		iii) the matrix $\ve j(\ve \theta^{\text{loc},i})$ defined in (\ref{eq:j}) is nonsingular.
		iv) weak signal condition, $\|\ve\theta^1-\ve\theta^0\|\leq c/\sqrt{L}$ with $c>0$.
Then, the asymptotic distribution of $\hat{\ve\theta}^\text{loc}$ under $\Hip_i$, $i=0,1$, is
\begin{align}\textstyle
\hat{\ve\theta}^\text{loc}\overset{a}{\sim}\N\left(\ve \theta^{\text{loc},i},\frac{1}{L} \ve j(\ve\theta^{\text{loc},i})^{-1}\tilde{\ve i}(\ve \theta^{\text{loc},i})\ve j(\ve\theta^{\text{loc},i})^{-1}\right),
\label{eq:lmle}
\end{align}
where the $(k,j)$ $M_k\!\times\! M_j$ sub-matrix of $\tilde{\ve i}(\ve \theta^{\text{loc},i})$ and $\ve j(\ve\theta^{\text{loc},i})$ are respectively defined by, $k,j\in[1:N]$, 
\begin{align} 
&\hspace*{-4pt}\textstyle [\tilde{\ve i}(\ve \theta^{\text{loc},i})]_{kj} \hspace*{-2.5pt}\equiv\Ex_{\ve\theta^i}\hspace*{-2pt}\left(\frac{\partial\log p_k(z_k(l);\ve\theta_k^\text{loc,i})} {\partial\ve\theta_k^\text{loc}} \frac{\partial\log p_j(z_j(l);\ve\theta_j^\text{loc,i})^T}{\partial\ve\theta_j^\text{loc}}\hspace*{-2pt}\right),
\label{eq:FisherLocal}\\
&\hspace*{-4pt}\textstyle [\ve j(\ve \theta^{\text{loc},i})]_{kj} \equiv -\Ex_{\ve\theta_k^{\text{loc},i}}\left( \frac{\partial^2 \log p_k(z_k(l);\ve\theta_k^\text{loc,i})} {\partial\ve\theta_k^\text{loc}\partial\ve\theta_j^\text{loc} } \right),\label{eq:j}
\end{align}
where the expectations are taken with respect to $p(\ve z_l;\ve\theta^{i})$, and the marginal PDF $p_k(z_k(l);\ve\theta_k^{\text{loc},i})$, respectively.
Also, the asymptotic distribution of $T_\text{L-MP}$ under $\Hip_i$ is:
\begin{align}
2\log T_{\text{L-MP}}(\ve z) & \overset{a}{\sim} f_P(\ve \mu_{ \text{MP},i},\ve{\Sigma}_{\text{MP},i}) \ i=0,1,
\label{eq:mpglr}
\end{align}
where $\ve\mu_{\text{MP},0}\!\!=\!\ve 0$, $\ve\mu_{\text{MP},1} = \sqrt{L} \ve i_\text{MP}(\ve\theta^\text{loc,0})^{\frac{1}{2}} (\ve\theta^{\text{loc},1}\!\!\!-\ve\theta^{\text{loc},0})$,~and
\noindent $\ve\Sigma_{\text{MP},i} \!\!=\! \ve i_\text{MP}(\ve\theta^{\text{loc},i})^{\frac{1}{2}} \ve j(\ve \theta^{\text{loc},i})^{-1} \tilde{\ve i}(\ve \theta^{\text{loc},i}) \ve j(\ve \theta^{\text{loc},i})^{-1}  \ve i_\text{MP}(\ve\theta^{\text{loc},i})^{\frac{1}{2}} $,
$$\textstyle \ve i_\text{MP}(\ve{\theta}^{\text{loc},i})\equiv \Ex_{\ve\theta^{\text{loc},i}}\Big(\frac{\partial\log p_\text{MP}(\ve z_l;\ve{\theta}^{\text{loc},i})}{\partial\ve{\theta}^\text{loc}} \frac{\partial\log p_\text{MP}(\ve z_l;\ve{\theta}^{\text{loc},i})^T}{\partial\ve{\theta}^\text{loc}} \Big),$$ 
with the expectation taken with respect to $p_\text{MP}(\ve z_l,\ve{\theta}^{\text{loc},i})$.
We also define\footnote{Note that $f_P(\ve 0_P, \mat I_P)$ ($f_P(\ve \mu, \mat I_P)$) is the (non-central) chi-2 PDF with $P$ degrees of freedom (with non-centrality parameter $\|\ve\mu\|^2$).} $f_P(\ve\mu,\ve\Sigma)$ as the PDF of $\|\ve n\|^2$ when $\ve n\sim\N(\ve\mu,\ve\Sigma)$.
\end{lemma}
In general, the PDF $f_P$ and its corresponding cumulative distribution function have not closed form expressions but can be tightly approximated using the Lugannani-Rice approximation \cite{lugannani1980saddle} given the fact that the corresponding moment generating function is easy to obtain. We will use this approximation for evaluating the miss-detection and false alarm probabilities of the statistic $T_\text{L-MP}$ for our numerical experiments presented in Sec. V.

For the rest of the paper, we will consider that the global vector parameter $\ve \theta$ can be expressed as $\ve \theta = \{\ve\theta_1,\dots,\ve\theta_N,\ve\theta^\text{ext} \} = \{\ve\theta^\text{loc}, \ve\theta^\text{ext}\}$, where $\ve\theta_k^\text{loc}=\ve\theta_k\in\mathbb{R}^{M_k}$, $k=1,\dots,N$ and $P=\sum_{k=1}^N M_k$ are the parameters observable locally at each node. Thus, local parameters at two different nodes do not have entries in common. Additionally, $\ve\theta^\text{ext}\in \R^{Q}$ ($M=P+Q$) are the parameters non-observable locally at any sensor, i.e., the parameters that do not modify the marginal distributions. This case is general enough, covers many problems of practical interest as we will see in the following sections, and allows a simpler treatment.

Next, we introduce the following corollaries, in which we analyze the so-called test against independence. In this test, we have exactly spatial independent measurements under the null hypothesis, i.e., $p(\ve z;\ve \theta^0) = \prod_{k=1}^{N}p_k(z_k;\ve \theta^{\text{loc},0}_k)$. However, under the alternative hypothesis, we have that the joint PDF is not necessarily the product of the marginal PDFs. The following corollary gives us simpler expressions than those in Lem. 1 for the asymptotic distributions of the L-MP statistic.
\begin{corollary}[Test against independence]
		Consider the assumptions of Lem. 1 and assume also that:  
		i) $p(\ve z_l;\ve \theta^0) = \prod_{k=1}^{N}p_k(z_k(l);\ve \theta^{\text{loc},0}_k)$
		Then, the asymptotic distribution of the L-MP statistic is
		\begin{equation}
		2\log T_\text{L-MP}(\ve z)\overset{a}{\sim}\left\{\begin{array}{ll}
		\chi^2_P & \text{under } \Hip_0\\
		\chi'^2_P(\|\ve \mu_{\text{MP},1}\|^2) & \text{under } \Hip_1,
		\end{array}\right.
		\label{eq:lmp2}
		\end{equation}
	\end{corollary}
	\begin{proof}
		In the following $\overset{a}{=}$ denotes equality when $L\rightarrow\infty$. We need to prove that, under the assumptions of the lemma, $\tilde{\ve i}(\ve\theta^\text{loc,1})\overset{a}{=} \tilde{\ve i}(\ve\theta^\text{loc,0})\overset{a}{=} \ve j(\ve\theta^\text{loc,1})\overset{a}{=}\ve j(\ve\theta^\text{loc,0})\overset{a}{=} \ve i_\text{MP}(\ve\theta^\text{loc,1})\overset{a}{=}\ve i_\text{MP}(\ve\theta^\text{loc,0})$, which implies that $\mat\Sigma_\text{MP,1} \overset{a}{=} \mat\Sigma_\text{MP,0} \overset{a}{=} \mat I_P$. Then, the result (\ref{eq:lmp2}) follows straightforwardly. We begin with the matrix $\tilde{\ve i}(\ve\theta^\text{loc,1})$ making a Taylor expansion of it around $\ve\theta^\text{loc,0}$. Using the weak signal assumption ($\ve\theta^\text{loc,1}\rightarrow\ve\theta^\text{loc,0}$ as $L\rightarrow\infty$), we get $\tilde{\ve i}(\ve\theta^\text{loc,1})\overset{a}{=} \tilde{\ve i}(\ve\theta^\text{loc,0})$. Then, using assumption i), we have that $\tilde{\ve i}(\ve\theta^\text{loc,0}) $ is a block diagonal matrix, i.e., $[\tilde{\ve i}(\ve\theta^\text{loc,0})]_{kj} = \Ex_{\ve\theta_k^{\text{loc},0}} \left( \frac{\partial \log p_k(z_k(l);\ve\theta_k^\text{loc,0})} {\partial\ve\theta_k^\text{loc}} \frac{\partial\log p_k(z_k(l); \ve\theta_k^\text{loc,0})^T} {\partial\ve\theta_k^\text{loc}} \right)\delta_{kj} = -\Ex_{\ve\theta_k^{\text{loc},0}}\left( \frac{\partial^2 \log p_k(z_k(l);\ve\theta_k^\text{loc,0})} {\partial\ve\theta_k^\text{loc}\partial\ve\theta_k^\text{loc} } \right)\delta_{kj}$.
		We proceed similarly (using a Taylor expansion and the weak signal assumption) to show that $\ve j(\ve\theta^\text{loc,1})\overset{a}{=}\ve j(\ve\theta^\text{loc,0})$ and $\ve i_\text{MP}(\ve\theta^\text{loc,1})\overset{a}{=}\ve i_\text{MP}(\ve\theta^\text{loc,0})$. Then, as $\ve\theta_k^\text{loc}$ and $\ve\theta_j^\text{loc}$, $k\neq j$, do not have parameters in common, we get $\ve j(\ve\theta^\text{loc,0})\overset{a}{=}\tilde{\ve i}(\ve\theta^\text{loc,0})$, and using assumption i) we get $\ve i_\text{MP}(\ve\theta^\text{loc,0})\overset{a}{=}\tilde{\ve i}(\ve\theta^\text{loc,0})$. Thus we get the result of the lemma.
\end{proof}

\subsection{Asymptotic deflection coefficients for GLR and L-MP}
The deflection coefficient of a statistic is commonly used as a simplified measure of performance when the error probabilities are difficult to compute. In the case that the statistic is Gaussian distributed \cite{Levy_Det}, the deflection coefficient characterizes completely the performance of a test. The deflection coefficient of $T$ is defined as $D_T = \frac{|\Ex_1(T)-\Ex_0(T)|} {\sqrt{\Var_0(T)}}$, where $\Var_0$ is the variance operator under $\Hip_0$. 
However, in some detectors where certain parameters are unknown and must be estimated through the data, even the mean or the variance can be rather cumbersome to compute. Therefore, we use the asymptotic characterization of the distributions to obtain closed-form expressions of the  deflection coefficients. For the GLR and L-MP the deflection coefficients can be written as:
\begin{align}
D_\text{G}= \frac{\lambda_g}{\sqrt{2 M}}, \, D_\text{L-MP} = \frac{\|\ve \mu_\text{MP,1}\|^2}{\sqrt{2}\|\mat\Sigma_{\text{MP},0}\|_F},
\label{eq:defcoefs}
\end{align} 
where we have use (\ref{eq:gglr}), Lem. 1, and that as $L\rightarrow\infty$, $\Ex_{\ve\theta^{\text{loc},1}}(2\log T_\text{L-MP}(\ve z))-\Ex_{\ve\theta^{\text{loc},0}}(2\log T_\text{L-MP}(\ve z)) = \|\ve \mu_{\text{MP},1}\|^2$, and
$\Var_0(2\log T_\text{L-MP}(\ve z)) =2\tr(\mat\Sigma_{\text{MP},0}^2)=2\|\mat\Sigma_{\text{MP},0}\|^2_F$, where $\|\cdot\|_F$ is the Frobenius norm\footnote{In the first equation we have neglected the trace term $\tr(\mat\Sigma_{\text{MP},1}-\mat\Sigma_{\text{MP},0})$ given that it does not depends on $L$, while $\|\ve \mu_{\text{MP},1}\|^2$ scales linearly with $L$.}.
Next corollary allows us to establish necessary and sufficient conditions under which the L-MP test performs better, worse or equal to the GLR test.
\begin{corollary}
		\label{cor:dc-comparison}
	Consider the assumptions of Corollary 1, and assume also that i) $\Ex_{\ve\theta^{0}}\left(\tfrac{\partial^2\log p(\ve z_l;\ve\theta^0)}{\partial\ve\theta_i\partial\ve\theta_j}\right)=\Ex_{\ve\theta_i^{\text{loc},0}}\left(\tfrac{\partial^2\log p_i(z_i(l); \ve\theta^\text{loc,0}_i)}{\partial\ve\theta_i^\text{loc}\partial\ve\theta_i^\text{loc}}\right)\delta_{ij}$, $i,j=1,\dots,N$. 
	Let $\lambda_\text{ext} = L(2(\ve\theta^\text{loc,1}-\ve\theta^\text{loc,0})^T \ve i_{12}(\ve\theta^0) (\ve\theta^\text{ext,1}-\ve\theta^\text{ext,0}) + (\ve\theta^\text{ext,1}-\ve\theta^\text{ext,0})^T \ve i_{22}(\ve\theta^0)(\ve\theta^\text{ext,1}-\ve\theta^\text{ext,0}) )$ and write the Fisher matrix in terms of $\ve i_{11}(\ve\theta^0)\in \R^{P\times P}$, $\ve i_{12}(\ve\theta^0)\in\R^{P\times Q}$ and 
	$\ve i_{22}(\ve\theta^0)\in \R^{Q\times Q}$ as 
	$$\ve i(\ve\theta^0) = \left[\begin{array}{ll}
	\ve i_{11}(\ve\theta^0)& \ve i_{12}(\ve\theta^0)\\
	\ve i_{12}(\ve\theta^0)^T& \ve i_{22}(\ve\theta^0)
	\end{array} \right].$$ 
	Then, $\ve i_{11}(\ve\theta^0) =\ve i_\text{MP}(\ve\theta^\text{loc,0})$, 
	\begin{equation}\textstyle
	D_\text{G}=D_\text{L-MP}\frac{\sqrt{P}}{\sqrt{M}}+ \frac{\lambda_\text{ext}}{\sqrt{2M}}\label{eq:DG-DLMP}
	\end{equation} 
	Moreover, $D_\text{L-MP}\geq D_\text{G}$ if and only if $D_\text{L-MP}\geq \frac{\lambda_\text{ext}}{\sqrt{2} (\sqrt{M}-\sqrt{P})}$.
\end{corollary}
\begin{proof}
	We begin showing that $\ve i_{11}(\ve\theta^0) =\ve i_\text{MP}(\ve\theta^\text{loc,0})$. Recall that $\ve\theta=\{\ve\theta^\text{loc},\ve\theta^\text{ext}\}$. Then, the $ij$-th block matrix $(\ve i_{11}(\ve\theta^0))_{ij}=-\Ex_{\ve\theta^0}\left(\frac{\partial^2\log p(\ve z_l;\ve\theta^0)}{\partial\ve\theta^\text{loc}_i\partial\ve\theta^\text{loc}_j}\right) = -\Ex_{\ve\theta_i^\text{loc,0}}\left(\frac{\partial^2\log p_i(z_i(l);\ve\theta^\text{loc,0})} {\partial\ve\theta^\text{loc}_i\partial\ve\theta^\text{loc}_i}\right)\delta_{ij}$, $i,j=1,\dots,N$, where we have used assumption i) in this corollary and the spatial independence of the data under $\Hip_0$. Then, using that the local parameters do not have entries in common, it is straightforward to show that $\ve i_{11}(\ve\theta^0) =\ve i_\text{MP}(\ve\theta^\text{loc,0})$. On the other hand, the GLR non-centrality parameter is $\lambda_g=L(\ve\theta^1-\ve\theta^0)^T\ve i(\ve\theta^0)(\ve\theta^1-\ve\theta^0) = \|\ve \mu_{\text{MP},1}\|^2 + \lambda_\text{ext}.$ Finally, using (\ref{eq:defcoefs}), and after some manipulations, we obtain the desired results.
\end{proof}

In general, the GLR test performance depends on a weighted sum (\ref{eq:DG-DLMP}) that considers the contribution of the local parameters (through $D_\text{L-MP}$) and the external parameters (through $\lambda_\text{ext})$. Then, this expression allows us to explain why in some scenarios the L-MP statistic outperforms the GLR test. Estimating all the $M$ parameters in the GLR test has the cost of increasing the overall variance of the statistic (cf. the denominator of the deflection coefficient), but not always improving the performance substantially through $\lambda_\text{ext}$. At the same time, if the number of external parameters increases, the quotient between $P$ and $M$ decreases, which also generates a decrease in $D_G$ according to eq. (\ref{eq:DG-DLMP}). In the special case in which there are not external or non-observable parameters i.e., $\ve\theta^\text{ext}=\emptyset$, $Q=0$, and $M=P$, Corollary \ref{cor:dc-comparison} implies that the L-MP and the GLR statistics are asymptotically equivalent, that is, they have the same asymptotic distribution (\ref{eq:lmp2}).
However, this is not the unique explanation for L-MP outperforming GLR in some settings. As we will see next in Sec \ref{sec:gaussian-3}, with correlated measurements under both hypotheses and where the amount of parameters to be estimated by both statistics is the same, we still find that the L-MP statistic outperforms the GLR test when the spatial correlation is not extremely high. 

\subsection{Gaussian model}
\label{sec:gaussian}
In this subsection we evaluate the theoretical deflection coefficients in (\ref{eq:defcoefs}) for both $T_\text{G}$ and $T_\text{L-MP}$ statistics for the following Gaussian test,
\begin{equation}
\Hip_0\!: \!{\ve z}_l \overset{iid}{\sim} \N(\ve \mu_0,\mat C_0 ),
\Hip_1\!\!: \!{\ve z}_l \overset{iid}{\sim} \N(\ve \mu_1,\mat C_1), l\in[1\!:\!L],
\label{eq:gaussian}
\end{equation}
where the parameters under $\Hip_0$ and $\Hip_1$ can be equal or different depending on the type of test to be done. We always assume that the parameters under $\Hip_0$ are known, except as otherwise indicated. We consider the following cases: the first case (test against independence) is presented as a case of study, and the second case will appear in the context of spectrum sensing in Sec. \ref{sec:SS}. In all the cases we use the well-known formula \cite[Sec. 3.9]{Kay_SSP_ET} for computing the corresponding Fisher information matrix for Gaussian distributions when needed.

\subsubsection{Case 1; test against independence}
\label{sec:gaussian-2}
In this case, $\ve\mu_0=\ve\mu_1=\ve 0$, $\mat C_0=\sigma^2_0 \mat I_N$ are known and $\mat C_1\neq\mat C_0$ is unknown. Without loss of generality, we take $\sigma^2_0=1$.
Then, the GLR parameter to be tested is $\ve\theta=\text{vech}(\mat C)\in R^M$, $M=\frac{1}{2}N(N+1)$, $\ve\theta^i=\text{vech}(\mat C_i)$, $i=0,1$. 
The deflection coefficient is given on the left side of (\ref{eq:dc}). On the other hand, the L-MP parameter to be tested is $\ve\theta^{\text{loc}}=\bar{\diag}(\mat C)\in \R^P$, $P=N$, $\ve\theta^{\text{loc},i}=\bar{\diag}(\mat C_i)$, while $\ve\theta^{\text{ext},i}\in\R^Q$ has as components all the sub-diagonal elements of $\mat C_i$, with $Q = \frac{1}{2}N(N-1)$, $i=0,1$. 
Using Corollary 1, it is straightforward to show that the L-MP deflection coefficient is (right side in the equation below):
\begin{equation}
D_\text{G}=\frac{L\|\mat C_1-\mat I_N\|^2_F}{2\sqrt{N(N+1)}}, \ D_\text{L-MP}=\frac{L\|\diag(\mat C_1)-\mat I_N\|^2_F}{2\sqrt{2N}}, 
\label{eq:dc}
\end{equation} 
In this case of testing against independence, we can apply Corollary \ref{cor:dc-comparison} to get the following necessary and sufficient condition to determine which of both statistics is better: $D_\text{L-MP}>D_\text{G}\Leftrightarrow$
$\|\diag(\mat C_1)-\mat I_N\|^2_F > \tfrac{2+\sqrt{2(N+1)}}{N-1} \|\text{offdiag}(\mat C_1)\|^2_F,$
where $\text{offdiag}(\mat C_1)$ is a matrix with the same off-diagonal elements of $\mat C_1$ but zeros in the main diagonal. As long as the difference of the local parameters at each node under $\Hip_1$ and $\Hip_0$ (i.e. the variances that appear in $\diag(\mat C_1)-\mat I_N$) be high enough compared to the contribution of the spatial correlation (determined by the off-diagonal elements of $\mat C_1$), the L-MP test will have a better deflection coefficient than the GLR test. 

\begin{figure}[t]
	\centering
	\includegraphics[width=1\linewidth]{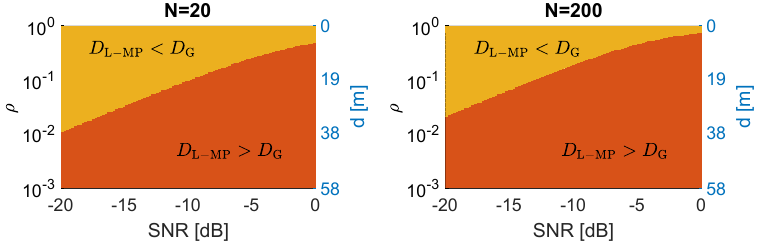}
	\caption{Test against the independence: comparison of L-MP and GLR statistics for an exponentially correlated model.}
	\label{fig:expocorr}
\end{figure}

Next, we consider an exponentially correlated model typically found in log-normal shadowing environments in which $(\mat C_1)_{ij}=\sigma_1^2\rho^{|i-j|}$, $\rho\in[0,1)$, $i,j=1,\dots,N$, and $\rho=e^{-d/d_0}$, where $d$ is the distance between two adjacent nodes and $d_0=8.33$m is the correlation distance for urban scenarios \cite{Gudmundson1991}. We define the SNR of the test as $\text{SNR}= \frac{\sigma^2_1-\sigma^2_0}{\sigma^2_0}$. 
In Fig. \ref{fig:expocorr}, we plot the parameter regions in the plane $(\text{SNR},\rho)$ for which $D_\text{L-MP}<D_\text{G}$ in yellow, and $D_\text{L-MP}>D_\text{G}$ in orange, for $N=20$ and $200$ nodes. We also added a second $y$-axis on the right of each figure showing the distance $d$ that corresponds to each $\rho$ shown in the left $y$-axis. As expected from (\ref{eq:dc}), we see that the L-MP test works better than the GLR test when the correlation level is lower than a given correlation threshold (the curve that divides the two regimes), which increases with the SNR. Note that, from a practical point of view, the distance between adjacent nodes for which the GLR test works better is relatively small, while, as we will see in the next subsection, the L-MP algorithm typically provides large communication overhead savings.  

\subsubsection{Case 2; correlation known shift-in-mean test}
\label{sec:gaussian-3}
We now consider the case in which $\ve\mu_0=\ve 0$, $\ve\mu_1\neq\ve 0$ is unknown and $\mat C_0=\mat C_1\equiv\mat C$ is known. As in general $\mat C_0\neq\mat I_N$, Corollary 1 and 2 do not apply to this example and Lem. 1 has to be used instead. 
The parameter test consists in testing the mean $\ve \theta =\ve\mu\in \R^N$, 
$\ve\theta^i =\ve \mu_i$, $i=0,1$,  with statistically dependent observations across the sensors.
In this case, $\ve\theta^\text{loc}=\ve \theta\in \R^N$, $M_k=1\ \forall k$, and $M=P=N$. 

\begin{figure}[t]
	\centering
	\includegraphics[width=1\linewidth]{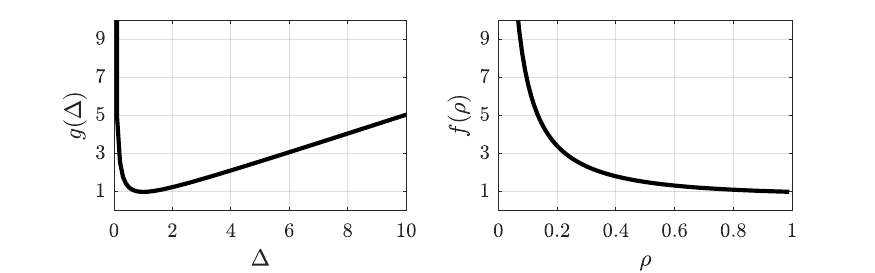}
	\caption{Functions to compare the deflection coefficients for $N=2$.}
	\label{fig:testmean}
\end{figure}

Using (\ref {eq:gglr}), the deflection coefficient of the GLR test
is shown on the left side of (\ref{eq:dc-mean}). On the other hand, using Lem. 1, the L-MP deflection coefficient is (right side in the equation below):
\begin{align}
D_\text{G} = \frac{L\ve\mu_1^T\mathbf{C}^{-1}\ve\mu_1}{\sqrt{2N}},  D_\text{L-MP} = \frac{L\ve\mu_1^T\diag(\mathbf{C})^{-1}\ve\mu_1}{\sqrt{2}\|\mat\Sigma_{\text{MP},0}\|_F},\label{eq:dc-mean}
\end{align}
where $\mat\Sigma_{\text{MP},0}=\diag(\mathbf{C})^{-1/2}\mathbf{C}\diag(\mathbf{C})^{-1/2}$. 
For a fixed $\mathbf{C}$, the behavior of the statistics depends on the variance in the direction of $\ve{\mu}_1$ given by $\mathbf{C}$ in the case of the GLR test, and $\diag(\mat C)$ for the L-MP test. The particular case of two nodes give us some intuition about the behavior of the statistics:  
Let $N=2$, $(\mat C)_{ij}\equiv\sigma_i\sigma_j\rho^{|i-j|}$, $\rho\in[0,1)$, $(\ve\mu_1)_i \equiv \mu_{1i}$, $i,j=1,2$, and $\Delta \equiv \frac{\sigma_2\mu_{12}}{\sigma_1\mu_{11}}$.  Then, evaluating (\ref{eq:dc-mean}), it is straightforward to show that: $D_\text{L-MP}\geq D_\text{G} \Leftrightarrow$
\begin{align}
g(\Delta)\equiv\tfrac{1}{2}(\Delta+ \Delta^{-1}) \leq \tfrac{\rho\sqrt{1+\rho^2}}{\sqrt{1+\rho^2}-(1-\rho^2)}\equiv f(\rho)
\label{eq:N2}
\end{align}
In Fig. \ref{fig:testmean}, we plot the functions previously defined against their corresponding arguments. It can proved that $f(\rho)$ is a monotonically decreasing function, $f(\rho)\rightarrow+\infty$ when $\rho\rightarrow 0$ and that $f(\rho)> 1$ $\forall\rho$. We see that for a given $\Delta=\Delta_0$, there exists a maximum correlation level $\rho_\text{max}\equiv f^{-1}(g(\Delta_0))$ such that $\rho\leq\rho_\text{max}$ $\Leftrightarrow D_\text{L-MP}\geq D_\text{G}$. That is, L-MP outperforms GLR if and only if the correlation coefficient is \emph{sufficiently} low. This is in line with the intuition about the L-MP statistic given that it is built using the product of the marginal PDFs. However, note that we can get $D_\text{L-MP}\geq D_\text{G}$ even for large values of $\rho$. For example, if $\Delta=1$, $g(1)=1$, and $D_\text{L-MP}\geq D_\text{G}$ $\forall\rho$.    
In Sec. \ref{sec:SS} we will see that the test analyzed in this subsection models a spectrum sensing scenario, and that the L-MP algorithm produces excellent results for scenarios typically found in practice.

\subsection{Communication resources analysis}
\label{sec:resources}
The communication between sensors for exchanging information requires that nodes access the wireless medium and it is one of the tasks that requires a substantial part of the energy budget of a WSN \cite{zhou2011modeling}. Therefore, in this subsection, we determine the amount of transmissions of a real scalar number\footnote{In practice, the amount of transmitted bits will depend on the quantization scheme.} per node needed to compute in a distributed fashion each algorithm for several scenarios.
Other energy consuming factors like computation tasks at each node are not tackled in this paper (although very relevant in practice) given that they strongly depend on the considered algorithm and also on the estimators to be computed. For example, if the local estimator has a closed-form expression, the computation complexity is typically small. However, if the estimator has to be computed with an iterative approach, the computation complexity is typically large.

In Table \ref{tab:cost}, we present all the statistics considered in the following analysis, including the case labeled as FRM (flooding raw measurements in the table) in which \emph{each} sensor measurement is communicated to the rest of nodes in the network using a flooding algorithm \cite{kshemkalyani2011distributed}. In this case, each node transmits $L N_f$ messages, where $N_f=O(|\mathcal{E}|)$. A fully-distributed algorithm designed for using minimal communication resources should employ substantially less transmissions than this strategy.

For the case of the L-MP algorithm, the communication cost is independent of the PDF of the data given that each summand of $\log T_\text{L-MP}$ in (\ref{eq:TLMP}) is estimated locally at each node without communicating with other nodes. Thus, the network needs to run only once the consensus algorithm to compute $\log T_\text{L-MP}$ as indicated in Algorithm \ref{fd}. Therefore, $N_{it}$ transmissions per node are required for computing this statistic. Note that average consensus can be achieved in a finite number of iterations if we employ the algorithm in \cite{sundaram_finite-time_2007}, where $N_{it}\leq N$.

For the case of the GLR test, the global MLE typically requires a large amount of sensor communications, even in the case when closed-form expressions are available. When this is not possible, numerical procedures are necessary, and the communication cost can be extremely high and may be impractical still under the assumption of spatial independent measurements \cite{blatt2004distributed}.  
The problem is more difficult to solve in the general scenario with spatially dependent measurements. In fact, to the best of our knowledge, a general procedure for efficiently obtaining a distributed MLE for models with general spatial dependent data is not available in the specialized literature. So we consider the Gaussian cases presented in the previous subsection (Sec. \ref{sec:gaussian-2} and \ref{sec:gaussian-3}), whose GLR statistics are given in Table \ref{tab:cost}, to compare the communication cost in each case.

In Case 1a (C.1a in the table), considering that $\mat C$ is the parameter of the GLR test, the sample covariance matrix must be estimated. The sample covariance matrix expression is not amenable to be implemented in a distributed scenario. However, we can use a \emph{suboptimal} approach given in \cite{wiesel2011distributed} for estimating the precision matrix (the inverse of the covariance matrix), assumed to be sparse (which is not the case in general). In this case, each node broadcasts its $L$ observations to its neighbors, and then estimates its corresponding row/column of the precision matrix (a maximum of $D$ coefficients can be estimated, where $D$ is the maximum degree of the network graph). Next, each node must communicate its estimation to the other nodes in the network using possibly a flooding algorithm transmitting $N_f D$ messages. Then, the total number of message exchanges is $L+N_f D$. 

Additionally, in Case 1a, it is clear that the GLR statistic can be computed in terms of the eigenvalues $\{\gamma_k\}$ of the empirical correlation matrix $\hat{\mat C}$ (C.1b). As we will see in Sec. V, some well-establish algorithms are based on all of them and others use only the maximum eigenvalue $\gamma_\text{max}$ (C.1c), which is a GLR test if the correlation matrix of the signal has rank 1 and the nodes have that additional information. This classification, using $\{\gamma_k\}$ or $\gamma_\text{max}$, is the only that matters for evaluating the communication cost, given that after computing the eigenvalue/s, the nodes can build the corresponding statistic as a function of it/them without extra communications. The communication cost can be computed as in \cite[Table II]{penna_decentralized_2015}, where in addition to an average consensus algorithm, it is necessary to iterate ($M_{it}$ times) a custom algorithm to compute the eigenvalues. Note that these algorithms are very expensive in terms of communication cost, in fact, they typically require more transmissions than the FRM strategy.

\begin{table}
	\centering
	\begin{tabular}{|c|l|c|l|l|}
		\hline 
		Algor.&Mdl. & Expression &\#comms. per node  &Expl.\\ 
		\hline 
		FRM& Gral. & Any & $L N_f$ & 1000\\
		\hline
		L-MP &Gral. & See (\ref{eq:TLMP}) & $N_{it}$ & 15\\
		\hline
		\parbox[t]{2mm}{\multirow{5}{*}{\rotatebox[origin=c]{90}{$2\log T_\text{G}/L$ }}}  
		&C.1a&  $-\log|\hat{\mat C}|+\tr(\hat{\mat C})$& $L + N_f D$ & 810\\ 
		&C.1b&  $\sum_{k=1}^N-\log\gamma_k+\gamma_k$   & $N_{it}(M_{it}L+M_{it})$ & 3300\\ 
		&C.1c&  $\gamma_\text{max}/\sigma_0^2$&  $N_{it}(M_{it}L+L+1)$ & 3165\\
		&C.2a&  $\hat{\ve\mu}^T\mat C^{-1}\hat{\ve\mu}$& $N_f$ & 100\\
		&C.2b&  $\hat{\ve\mu}^T\hat{\mat C}^{-1}\hat{\ve\mu}$&  $L+N_f (D+1)$ & 810\\
		\hline 
	\end{tabular} 
	\caption{\small Number of transmissions for transmitting all the measurements (FRM), and computing the L-MP, and the GLR statistics. 
As an example, consider $N=40$, $|\mathcal{E}|=100$, $N_{it}=15$, $N_f=100$, $L=10$, $D=8$ and $M_{it}=20$.}
	\label{tab:cost}
\end{table}

In Case 2a (C.2a), $\mathbf{C}$ is assumed to be known, so each node can compute its corresponding sample mean $(\hat{\ve \mu})_k$ and communicate it to the rest of the nodes using the flooding algorithm with a cost of $N_f$ messages per node. Finally, in Case 2b (C.2b), $\mathbf{C}$ is unknown, so we have to add the cost of estimating the sample covariance matrix $\hat{\mat C}$ resulting in $L+N_f(D+1)$ messages.

In Table \ref{tab:cost}, we also evaluate the communication cost for the same scenario considered in \cite{penna_decentralized_2015} (see Expl. column). As it is shown, the implementation of the GLR statistic requires in general elevated communication network resources. Moreover, when a test based on either the maximum eigenvalue or the eigenvalues is considered, the network resources are  extremely high, even larger than transmitting all the measurements of each sensor to the others. This is a notorious advantage in favor of $T_\text{L-MP}$ when the achieved performance is satisfactory.

\section{Application to spectrum sensing}
\label{sec:SS}

In this section we consider as an application example a cognitive radio (CR) system, which emerged several years ago as a possible solution for the spectrum shortage (see \cite{2015PoorSpectrumSensing} 
and references therein). 
In CR systems, unlicensed, or secondary users (SUs) sense the spectrum in a particular place and time and wish to detect the presence or absence of licensed, or primary users (PUs), in order to use the spectrum when it is available.
Consider a CR system where the $k$-th SU, $k=1,\dots,N$, is located at a distance $d_k$ from the primary transmitter (PTx). The secondary network may opportunistically access the PUs band if $d_k$ is greater than a certain length\footnote{This distance depends on several parameters such as the transmitted powers of the primary and the secondary transmitters, the maximum interference tolerated by the primary receivers and the coverage range of the primary network.}, $R_p$, defined as the primary range. Thus, the SUs has to determine if $d_k = R_p + c_k$ for some $c_k > 0$, $k=1,\dots,N$.
Assume that the $k$-th SU measures the signal-to-noise ratio (SNR) of the primary signal, on a logarithmic scale. In practice, the $k$-th sensor could measure the received power from the PTx and compute its corresponding SNR relative to the noise floor in the receiver.
Then, under log-normal shadowing and assuming perfect SNR estimation, the sensor measurement has a Gaussian distribution with mean $\text{PL}_\text{dB}(d_k)$ determined by the distance-dependent path-loss. Let $z_k(l)$ be the normalized SNR measurement of the $k$-th SU at the $l$-th time slot, obtained by subtracting to the original measurement the mean SNR obtained at $R_p$. Thus, the cooperative spectrum sensing problem may be formulated as the following binary hypothesis testing problem \cite{ghasemi2007asymptotic}, 
\begin{equation}
\!\Hip_0: {\ve z}_l \overset{iid}{\sim} \N(\ve 0,\mat C ),\ 
\Hip_1: {\ve z}_l \overset{iid}{\sim} \N(\ve \mu_1,\mat C), \ l\in[1:L],
\label{eq:gaussiantest}
\end{equation}
where the $k$-th component of $\ve \mu_1$ is $\mu_{1k}=\text{PL}_\text{dB}(d_k)-\text{PL}_\text{dB}(R_p)$, $k=1,\dots,N$, and $\mat C$ is the covariance matrix determined by the shadowing effect. Assuming that the sensing network is
small compared to the primary range, it can be considered that $\mat C$ is independent of the distance $d_k$. Following the experimental evidence provided in \cite{Gudmundson1991}, we consider i) $\text{PL}_\text{dB}(d_k)= K-10\alpha\log_{10}(d_k)$, where $K$ is the attenuation constant (in dB), $\alpha$ is the path-loss exponent and the distance $d_k$ is in meters, and ii) an exponential decreasing function for the correlation model, i.e., $(\mat C)_{ij}=\sigma^2_\text{sh} e^{-\frac{d_{ij}}{d_0}}$, where $d_{ij}$ is the distance between SU $i$ and SU $j$, $d_0$ is the correlation distance, and $\sigma^2_\text{sh}$ is the common variance at each SU site. The last two parameters depend on the shadowing environment (e.g., an urban or suburban scenario). 
Table \ref{tab:sh} summarizes the parameters for the considered shadowing scenarios obtained from \cite{Gudmundson1991}.
\begin{figure}[t]
	\centering
	\includegraphics[width=1\linewidth]{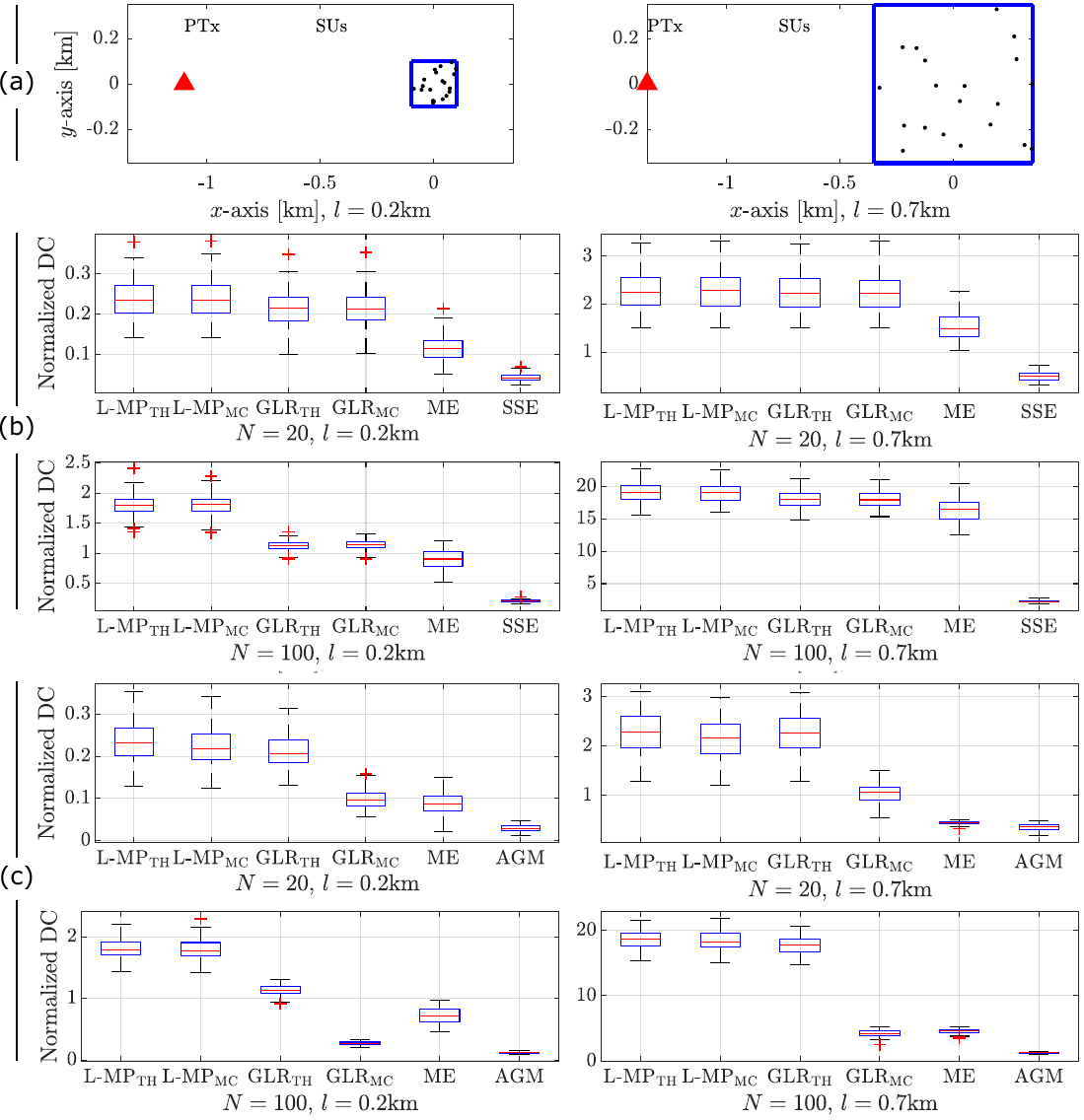}
	\caption{Urban scenario: (a) Top line: Network configuration. Box-plots of the normalized DC for $100$ network realizations with (b) $\mat C$ known, and (c) $\mat C$ unknown.}
	\label{fig:bp_urb}
\end{figure}
\begin{figure}[h]
	\centering
	\includegraphics[width=1\linewidth]{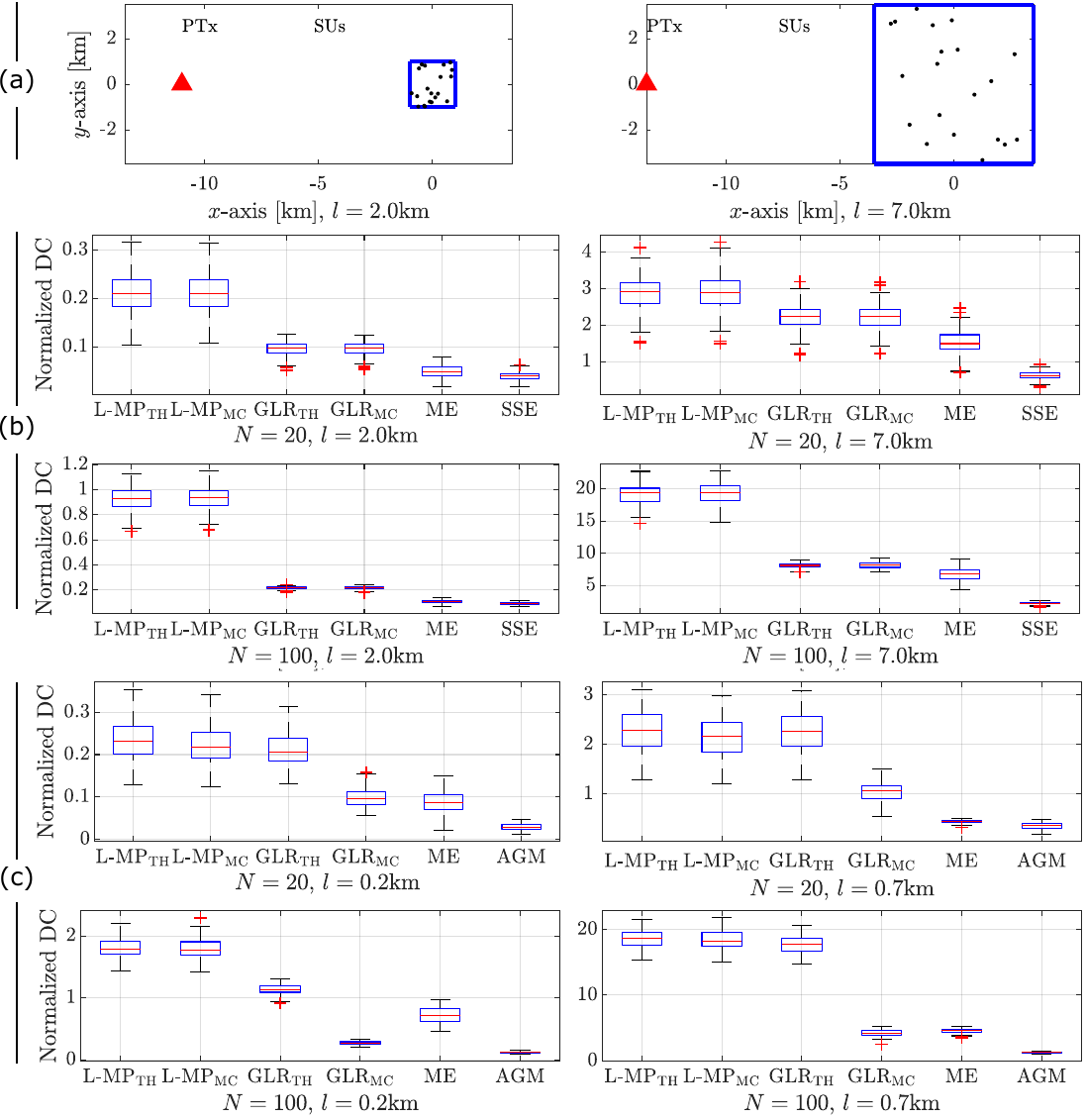}
	\caption{Suburban scenario: (a) Top line: Network configuration. Box-plots of the normalized DC for $100$ network realizations with (b) $\mat C$ known, and (c) $\mat C$ unknown.}
	\label{fig:bp_suburb}
\end{figure}
\begin{figure}
	\centering
	\includegraphics[width=\linewidth]{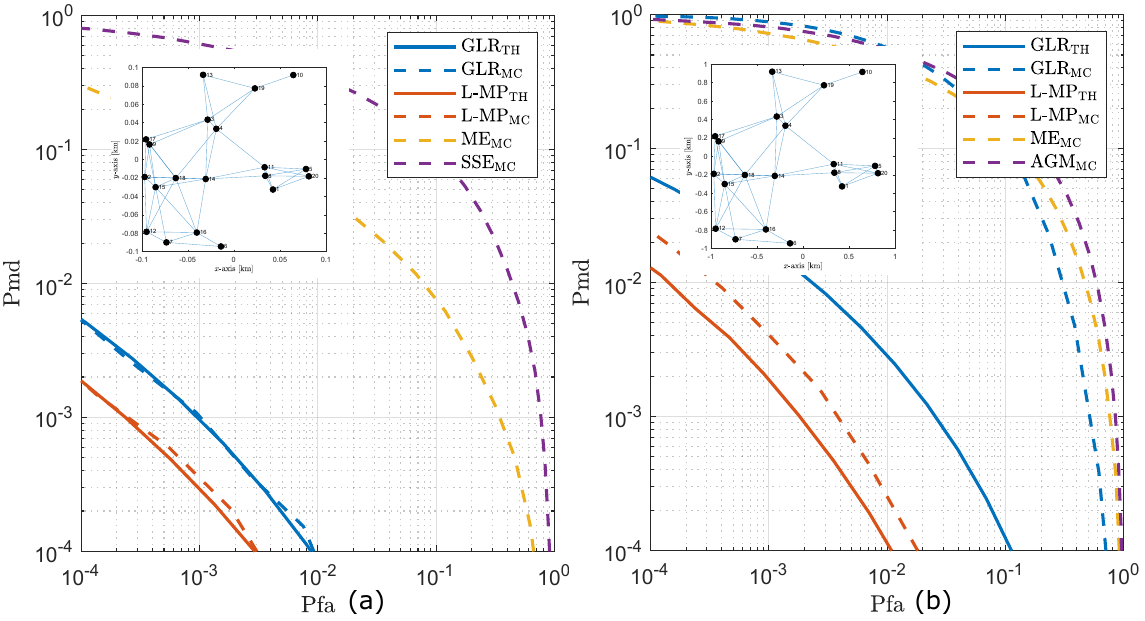}
	\caption{CROCs for the network realizations shown in each picture: (a) Urban scenario with $\mat C$ known, $N=20$, $L=21$, and $l=0.2$km. (b) Suburban scenario with $\mat C$ unknown, $N=20$, $L=30$, and $l=2$km.}
	\label{fig:rocs}
\end{figure}
\begin{table}[t]
	\centering 
	\begin{tabular}{|c|c|c|c|c|}
		\hline 
		Scenario & Freq. band [MHz]  &$\alpha$  &$\sigma_\text{sh}$ [dB] &$d_0$ [m]	\\ 
		\hline 
		Urban    & 1700      		 &5   	 	&4.3   					 &8.33  	\\ 
		\hline
		Suburban & 900  			 &3.3  		&7.5  					 &500  		\\ 
		\hline 
	\end{tabular} 
	\caption{Parameters for the considered shadowing scenarios. 
	}
	\label{tab:sh}
\end{table}
\begin{table}[b]
	\centering
	\begin{tabular}{|l|l|c|}
		\hline 
		Case & Statistic expression & Ref.\\ 
		\hline 
		$\mat C$ known	&$2\log T_\text{L-MP}/L=\hat{\ve\mu}^T\diag(\mat C)^{-1}\hat{\ve\mu}$& See (\ref{eq:TLMP})\\ 
		&$2\log T_\text{G}/L=\hat{\ve\mu}^T\mat C^{-1}\hat{\ve\mu}$& See above (\ref{eq:gglr})\\
		&$T_\text{ME}=\gamma_\text{max}/\sigma^2_\text{SH}$& \cite{zeng2008blindly,taherpour_multiple_2010} \\
		&$T_\text{SSE}= \sum_{k=1}^N -\log(1+\beta_k)+\beta_k$&\cite{zhang_multi-antenna_2010}\\
		\hline 
		\hline
		$\mat C$ unknown&$2\log T_\text{L-MP}/L=\hat{\ve\mu}^T\diag(\hat{\mat C})^{-1} \hat{\ve\mu}$& See (\ref{eq:TLMP})\\
		&$2\log T_\text{G}/L=\hat{\ve\mu}^T\hat{\mat C}^{-1}\hat{\ve\mu}$& See above (\ref{eq:gglr})\\
		&$T_\text{ME}=\gamma_\text{max}/(\sum_{k=1}^N\gamma_k)$&\cite{taherpour_multiple_2010,bianchi_performance_2011}\\
		&$T_\text{AGM}= (\tfrac{1}{N}\sum_{k=1}^N \gamma_k)/(\prod_{k=1}^N\gamma_k)^\frac{1}{N}$ & \cite{zhang_multi-antenna_2010}\\
		\hline		
	\end{tabular}
	\caption{\small Statistics expressions for the test (\ref{eq:gaussiantest}), where 
		$\gamma_k$ is the $k$-th eigenvalue of the sample correlation matrix $\hat{\mat R}= \frac{1}{L}\sum_{l}\ve z_l\ve z_l^T$, $\gamma_\text{max}=\max_k\gamma_k$, $\beta_k\equiv\max(\gamma_k/\sigma^2_\text{SH}-1,0)$, $k=1,\dots,N$, $\hat{\mat C} = \frac{1}{L}\sum_{l}(\ve z_l-\hat{\ve\mu})(\ve z_l-\hat{\ve\mu})^T$, and $\hat{\ve \mu} = \frac{1}{L}\sum_{l}\ve z_l$.}
	\label{tab:stats}
\end{table}
In Figs. \ref{fig:bp_urb}(a) and \ref{fig:bp_suburb}(a) (top two plots of each figure) we see a schematic representation of the network for an urban and suburban scenario, respectively. We consider that the SUs are uniformly distributed in a square (in blue) of side $l$, and that the distance from its nearest side to the PTx is $R_p$. Let $l=0.2$km (left column) and $l=0.7$km (right column), and $R_p=1$km for the urban scenarios; and $l=2$km (left column) and $l=7$km (right column), and $R_p=10$km for the suburban scenarios. The black dots represent a realization of the SUs location. 

In the following we consider that $\mat C$ can be known or unknown. In the case that $\mat C$ is unknown, it is a nuisance parameter \cite{Kay_SSP} and must be estimated (in the case of the L-MP only the elements of the diagonal of $\mat C$ are necessary). It can be proved that the asymptotic performance of both the L-MP and the GLR statistics is exactly the same when $\mat C$ is known or unknown. However, this is not true in the finite length regime. The statistics for the GLR test and the L-MP test are given in Table \ref{tab:stats}. We take $L>N$ in each scenario in order for the sample covariance matrix to be non-singular with probability 1. Thus, for each case we consider i) $N=20$ and $L=30$, and ii) $N=100$ and $L=110$.

For comparison purposes, we also consider the statistics shown in Table \ref{tab:stats} based on the eigenvalues of the sample correlation matrix typically used in the context of spectrum sensing, which inherently take into account the spatial correlation of the node measurements.  
As the measure of performance we first report the normalized deflection coefficient (DC) of a given statistic $T$, i.e., $D_T/L$, which of course depends on the SUs positions. Thus, for each network realization, we have a different DC, and therefore we show the corresponding box-plot which informs the median (in red), the first and third quartiles and the outliers of $D_T/L$ for each scenario. Note that the detection performance depends on the unnormalized deflection coefficient $D_T$, so, a small difference in $D_T/L$ generally implies a large difference in performance if $L$ is large.

In the case of the GLR and L-MP statistics we have the theoretical expressions in (\ref{eq:dc-mean}), labeled GLR$_\text{TH}$ and L-MP$_\text{TH}$ respectively, computed using the asymptotic distributions. In order to evaluate their behavior in the finite length regime, we also compute the DCs of the statistics using the method of Monte Carlo with $10^3$ trials, labeled GLR$_\text{MC}$ and L-MP$_\text{MC}$, respectively. On the other hand, the DCs of the remaining statistics (Maximum Eigenvalue, ME; Subspace Eigenvalue, SSE; and Average to Geometric Mean, AGM) are also computed using the method of Monte Carlo with the same amount of trials.  
  
In Fig. \ref{fig:bp_urb}(b), we have the urban scenario with $\mat C$ known for the indicated number of SUs $N$, and square side $l$. First, we see that the asymptotic statistic (L-MP$_\text{TH}$ and GLR$_\text{TH}$) are in agreement with their finite counterparts (L-MP$_\text{MC}$ and GLR$_\text{MC}$) for both statistics. We also see that the L-MP statistic outperforms the other statistics, something that is more evident for $N=100$ and $l=0.2$km. We also remark that this behavior occurs with high probability in terms of spatial distributions of the SUs, i.e., for about $99.9$\% of the network realizations the L-MP outperforms the GLR statistic, which has the second best performance for this scenario.  
Note that when $\mat C$ is known, the L-MP and the GLR statistic estimate the same amount of parameters ($N$) for testing (\ref{eq:gaussiantest}): both estimate the vector mean. Therefore, the difference of performance can be explained through the different structure of the statistic: while the L-MP uses $\diag(\mat C)$ in the quadratic form, the GLR uses $\mat C$. Recalling the case with two nodes (c.f. (\ref{eq:N2})), if a large amount of the $N$ nodes have similar means (small distance between nodes) and variance (it is actually the same for all nodes $\sigma^2_\text{SH}$), the quantity $\Delta$ is near to $1$ for each pair of nodes. Thus, the L-MP works better than the GLR for almost any correlation level, even when the correlation is high due to the distance between two nodes is small. This observation could explain why the L-MP has a DC with more \emph{relative} difference with respect to the GLR when the nodes are close to each other (i.e., $l=0.2$km).

Regarding the remaining statistics for this scenario (ME and SSE), their poor performance is due to the fact that they were developed for other scenarios, where the measurements are uncorrelated under $\Hip_0$, which is not the case here. We have included them here only because they are typically considered in spectrum sensing applications.

In Fig. \ref{fig:rocs}(a), we present the complementary receiver operating characteristics (CROCs) of the statistics for a realization of the network (see inside of the figure) for the same urban scenario with $\mat C$ known and the parameters shown in the caption of the figure. Note that the performance of the algorithm is in agreement with the behavior predicted by the deflection coefficients: we see that the L-MP test outperforms the GLR test and that the theoretical error probabilities computed in the asymptotic scenario for both statistics (labeled GLR$_\text{TH}$ and L-MP$_\text{TH}$) present an excellent agreement with the error probabilities computed using the method of Monte Carlo for the finite length regime (labeled GLR$_\text{MC}$ and L-MP$_\text{MC}$). This is due to the distribution of the estimated vector parameter (the mean) conditioned to each hypothesis in the  finite regime is Gaussian, which is indeed its asymptotic distribution. 

In Fig. \ref{fig:bp_urb}(c), overall we have similar conclusions than before for the same urban scenario, but when $\mat C$ is unknown. Note however that the DCs of the GLR computed theoretically and using Monte Carlo present more discrepancies than before. This behavior will be explained next. Now we remark that the results obtained by the L-MP test with $\mat C$ unknown are important from a practical point of view, given that in order to know $\mat C$, each node should know precisely the position of each SU of the network, which could be non-realistic.

In Fig. \ref{fig:bp_suburb}(b) and (c) we show similar results now for the suburban scenario, for which the correlation is relatively bigger than in the urban scenario for the considered network configuration: while the size of the secondary network increases 10 times (from $l=0.7$km to $l=7$km), the correlation distance $d_0$ increases about 60 times (from $8.33$m to $500$m). 
If we look at the DCs and the CROCs of the GLR test in Fig. \ref{fig:rocs}(b) with $\mat C$ unknown (i.e., a nuisance parameter) for the same suburban scenario with the parameters shown in the caption of the figure, we observe that the GLR test also presents discrepancies between the theoretical and the Monte Carlo performances (compare GLR$_\text{TH}$ with GLR$_\text{MC}$). When $\mat C$ is unknown, the GLR has to estimate also the complete covariance matrix and, thus, the distribution of the nuisance vector parameter is strictly Gaussian only in the asymptotic scenario (using the central limit theorem), when $L\rightarrow\infty$. So, the considered $L=30$ is too low to be even near the asymptotic scenario. Note however that the theoretical and Monte Carlo CROCs of the L-MP statistic present a much better agreement than the GLR test. This can be attributed to the fact that the L-MP test has to estimate much less nuisance parameters ($N$, the diagonal elements of the covariance matrix) than the GLR test ($\frac{1}{2}N(N+1)$).  

Finally, note that for all the configurations presented ($\mat C$ known/unknown, urban/suburban scenario and $N=20/100$), the performance of all algorithms is increased when the same number of nodes are distributed in a larger area, i.e., they are allowed to be more distant one from the other (see Figs. \ref{fig:bp_urb} and \ref{fig:bp_suburb} and compare the performance of the left column against the one of the right column). This is in concordance with the results reported in \cite{ghasemi2007asymptotic} for nodes distributed uniformly in a line using the same correlation model.

\section{Concluding remarks}
In this paper we have characterized the asymptotic performance of a fully-distributed detection algorithm for a WSN with observations following a general parametric distribution. The proposed statistic successfully reduces the amount of local transmissions to build the statistic by estimating the unknown parameters locally, and thus, saving valuable network resources as energy, bandwidth or delay.
The analytical and numerical results show that in some scenarios with data distribution with unknown parameters, the advantages of using  $T_\text{L-MP}$ is twofold: it performs better than commonly-used statistics and its implementation is much more efficient in terms of network resources. That is, in some cases, the fact of implementing this simpler strategy 
not only do not introduce penalties but can also be beneficial in composite tests. 
In general, the tools developed in this paper allow to quantitatively characterize the improvement/penalty introduced by not using the full, but also more complex, parametric structure of the measurements distribution. In those scenarios for which the L-MP introduces significant penalties, future research directions could include investigating methods that take into account, at least partially, the dependence structure of the data for building a fully-distributed detection algorithm. For example, the variational Bayes and graph signal processing techniques could be good candidates for this task.

\iftrue
\appendices
\section{Proof of Lemma 1}
The proof of the lemma can be splitted in two parts: the asymptotic distribution of the local MLE,
and the asymptotic distribution of the statistic $T_\text{L-MP}$. We will refer to $\ve\theta^\text{loc*}$ as the true parameter of the distribution ($\ve\theta^\text{loc*} = \ve\theta^{\text{loc},0}$ under $\Hip_0$ and $\ve\theta{^\text{loc*}} = \ve\theta^{\text{loc},1}$ under $\Hip_1$).
Before starting with the proof we need to present the following result:
\begin{theorem} [Mean value theorem] \cite[Th. 12.9]{apostol1974mathematical}
	Let $B$ be an open subset of $\R^P$ and assume that $\ve f: B \rightarrow \R^P$ is differentiable at each point of $B$. Let $\ve x$ and $\ve y$ be two points in $B$ such that the segment $S(\ve x, \ve y)= \{t\ve x+ (1-t)\ve y: t\in [0,1] \} \in B$. Then for every vector $\ve a$ in $\R^P$ there is a point $\ve w\in S(\ve x, \ve y)$ such that
	$\ve a^T (\ve f(\ve y) - \ve f(\ve x)) = \ve a^T J(\ve w)(\ve y - \ve x)$,
	where $J(\ve w)$ is the Jacobian matrix of $\ve f$ evaluated in $\ve w$, i.e., $[J]_{i,j}=\frac{\partial f_i}{\partial x_j}$, where $\{f_i\}$ are the components of $\ve f$. It is important to remark that $\ve w$ depends on $\ve a$.
	\label{theo:mean}
\end{theorem}
\subsection{Asymptotic distribution of the local MLE}
\label{app:lem1-a}

Given that the hypothesis $\Hip_0$ is simple ($\ve\theta^{\text{loc},0}$ is a known vector), we have to consider the unrestricted MLE. Let 
$\ve \psi(\ve z_l;\ve\theta^\text{loc})\in\R^{P}$ be $\ve \psi(\ve z_l;\ve\theta^\text{loc})= \left[\frac{\partial\log p_1(z_1(l); \ve\theta_1^\text{loc})^T}{\partial\ve\theta_1^\text{loc}},\dots,\frac{\partial\log p_N(z_N(l); \ve\theta_N^\text{loc})^T}{\partial\ve \theta_N^\text{loc}}\right]^T.$
By definition, the MLE must satisfy 
\begin{equation}
\textstyle \frac{1}{L}\sum_{l=1}^{L} \ve\psi(\ve z_l;\hat{\ve\theta}^\text{loc})=\ve 0.
\label{eq:MLE_deriv}
\end{equation}
Consider Theorem \ref{theo:mean} with $\ve f(\ve \theta^\text{loc}) = \frac{1}{L}\sum_{l=1}^{L} \ve\psi(\ve z_l;\ve\theta^\text{loc})$, $\ve x = \ve\theta^\text{loc*}$ and $\ve y =\hat{\ve\theta}^\text{loc}$, then we have
$\ve a^T ( \frac{1}{L}\sum_{l=1}^{L} \ve\psi(\ve z_l;\hat{\ve\theta}^\text{loc})- \frac{1}{L}\sum_{l=1}^{L} \ve\psi(\ve z_l;\ve\theta^\text{loc*}))
= \ve a^T J(\ve w^L)(\hat{\ve\theta}^\text{loc}-\ve\theta^\text{loc*})$ $\forall \ve a\in\mathbb{R}^N$, 
where $\ve w^L$ belongs to the segment $S(\hat{\ve\theta}^\text{loc},\ve\theta^\text{loc*})$ and depends on $\ve a$. 
Using (\ref{eq:MLE_deriv}) in the previous equation produces
\begin{align}
\textstyle  \ve a^T \left\{\frac{1}{\sqrt{L}}J(\ve w^L)\left( J(\ve w^L)^{-1} \frac{1}{\sqrt{L}}\textstyle\sum_{l=1}^{L} \ve\psi(\ve z_l;\ve\theta^\text{loc*})\right.\right. \nonumber\\
\left.\left.+ \sqrt{L}(\hat{\ve\theta}^\text{loc}-\ve\theta^\text{loc*})\right)\right\}=0, \ \ \forall \ve a.
\label{eq:mvt}
\end{align}  
By consistency of the estimator $\hat{\ve\theta}^\text{loc}$, the segment $S(\hat{\ve\theta}^\text{loc},\ve\theta^\text{loc*})$ becomes the point $\ve\theta^\text{loc*}$ and 
$\ve w^L\overset{p}{\rightarrow}\ve\theta^\text{loc*}$ as $L\rightarrow\infty$. 
Thus, the expression inside the parenthesis in (\ref{eq:mvt}) becomes independent of $\ve a$ as $L\rightarrow\infty$, and therefore, it must converge in probability to $\ve 0$. Then,
using the continuity of the second-order partial derivatives of the log-likelihood function, we apply the Continuous Mapping Theorem (CMT)\cite{Vaart_2000} to obtain $J(\ve w^L)\overset{p}{\rightarrow}-\ve j(\ve\theta^\text{loc*})$, where $\ve j(\ve{\theta}^\text{loc*})$ is defined in (\ref{eq:j}).
Additionally, by the multivariate central limit theorem (CLT)
\begin{equation*}
\textstyle \frac{1}{\sqrt{L}}\sum_{l=1}^{L} \ve\psi(\ve z_l;\ve\theta^\text{loc*}) \overset{a}{\sim} \N(\ve 0,\tilde{\ve i}(\ve \theta^\text{loc*})),
\end{equation*}
where the mean of the Gaussian distribution is $\ve 0$ by the first assumption and its covariance matrix is defined in (\ref{eq:FisherLocal}).
Finally, 
\begin{align*}
\textstyle \sqrt{L}(\hat{\ve\theta}^\text{loc}-\ve\theta^\text{loc*})\overset{a}{\sim} 
-J(\ve w^L)^{-1} \frac{1}{\sqrt{L}}\sum_{l=1}^{L} \ve\psi(\ve z_l;\ve\theta^\text{loc*}) \nonumber \\ \overset{a}{\sim} \N(\ve 0,\ve j(\ve{\theta}^\text{loc*})^{-1}\tilde{\ve i}(\ve \theta^\text{loc*})\ve j(\ve{\theta}^\text{loc*})^{-1} )
\end{align*}  
Solving for $\hat{\ve\theta}^\text{loc}$ we obtain the first result of the lemma in (\ref{eq:lmle}).
\subsection{Asymptotic distribution of $T_\text{L-MP}$}
\label{app:lem3}
To prove this lemma we need fundamentally to show that the following factorization is valid:  
\begin{equation}
\textstyle \frac{\partial\log p_\text{MP}(\ve z_l;\ve{\theta}^\text{loc*})}{\partial\ve{\theta^\text{loc}}} = \ve i_\text{MP}(\ve{\theta}^\text{loc*})(\hat{\ve{\theta}}^\text{loc}-\ve{\theta}^\text{loc*}),
\label{eq:facMP}
\end{equation}
where $\hat{\ve{\theta}}^\text{loc}$ is the local MLE estimator. Note that this is a similar factorization to that one found for estimators attaining the Cramer-Rao bound but with the true joint PDF replaced by $p_\text{MP}(\cdot,\ve{\theta}^\text{loc*})$. We show next that this equation is valid even when $\hat{\ve{\theta}}^\text{loc}$ does not attain, in general, the Cramer-Rao bound, asymptotically achieved by the global MLE.

Hereafter, where there is no ambiguity, we will drop the supra index $\text{loc}$ from $\hat{\ve{\theta}}^\text{loc}$ and $\ve{\theta}^\text{loc*}$ and call them $\hat{\ve{\theta}}$ and $\ve{\theta}$, respectively, in order to shrink the size of the equations.
Given that the local MLE $\hat{\ve{\theta}}$ is consistent, $\ve{\theta}=\Ex_{\ve{\theta}^\text{loc*}}(\hat{\ve{\theta}}) = \int \hat{\ve\theta} p_\text{MP}(\ve z;\ve{\theta})d\ve z$ is asymptotically satisfied when $L\rightarrow\infty$. Then
\begin{align*}
\textstyle\frac{\partial\ve{\theta}}{\partial\ve{\theta}} = \mathbf{I}_P &= \textstyle\int \hat{\ve{\theta}} \frac{\partial p_\text{MP}(\ve z;\ve{\theta})^T}{\partial\ve{\theta}}d\ve z\\
& =\textstyle \int \hat{\ve{\theta}} \frac{\partial\log p_\text{MP}(\ve z;\ve{\theta})^T}{\partial\ve{\theta}}p_\text{MP}(\ve z;\ve{\theta})d\ve z\\
& = \textstyle\int (\hat{\ve{\theta}}-\ve{\theta}) \frac{\partial\log p_\text{MP}(\ve z;\ve{\theta})^T}{\partial\ve{\theta}}p_\text{MP}(\ve z;\ve{\theta})d\ve z
\end{align*}
where in the last equality we used the second assumption. Let $\ve a,\ve b\in\R^P$ arbitrary vectors. After pre- and post-multiplication of the last equation by $\ve a^T$ and $\ve b$, respectively, we have:
\begin{equation}
\textstyle\ve a^T\ve b = \int \ve a^T (\hat{\ve{\theta}}-\ve{\theta}) \frac{\partial\log p_\text{MP}(\ve z;\ve{\theta})^T}{\partial\ve{\theta}}\ve b \ p_\text{MP}(\ve z;\ve{\theta})d\ve z
.\label{eq:int}
\end{equation}

Then, we need the following Cauchy-Schwarz inequality 
$$\left[\int w(\ve z)g(\ve z)h(\ve z)d\ve z \right]^2\leq \int w(\ve z)g^2(\ve z)d\ve z\int w(\ve z) h^2(\ve z)d\ve z, $$
where $g(\ve z)$ and $h(\ve z)$ are arbitrary scalar functions and $w(\ve z)\geq 0, \forall\ve z$, and the equality holds if and only if $g(\ve z)=c\, h(\ve z)$.  Let $w(\ve z)=p_\text{MP}(\ve z;\ve \theta)$, $g(\ve z)=\ve a^T(\hat{\ve{\theta}}-\ve{\theta})$ and $h(\ve z)=\frac{\partial\log p_\text{MP}(\ve z;\ve{\theta})^T}{\partial\ve{\theta}}\ve b$ and apply the Cauchy-Schwarz inequality to (\ref{eq:int}) to obtain
\begin{align}
\textstyle(\ve a^T\ve b)^2&\textstyle\leq \int \ve a^T(\hat{\ve{\theta}}-\ve\theta)(\hat{\ve{\theta}}-\ve\theta)^T\ve a \ p_\text{MP}(\ve z,\ve \theta)d\ve z\label{eq:cs0}\\
& \textstyle\times\int \ve b \frac{\partial\log p_\text{MP}(\ve z;\ve{\theta})}{\partial\ve{\theta}}\frac{\partial\log p_\text{MP}(\ve z;\ve{\theta})^T}{\partial\ve{\theta}}\ve b \ p_\text{MP}(\ve z,\ve \theta)d\ve z\nonumber\\
&\textstyle {=} \ve a^T \mathbf{C}_\text{MP}\ve a \ \ve b^T \ve i_\text{MP}(\ve\theta)\ve b \label{eq:cs1}\\
& \textstyle{=} \ve a^T \mathbf{C}_\text{MP}\ve a \ \ve a^T\ve b\label{eq:cs2}
\end{align}
where in (\ref{eq:cs1}) we defined $\mathbf{C}_\text{MP} = \Ex_{\ve{\theta}^\text{loc*}}((\hat{\ve{\theta}}-\ve\theta)(\hat{\ve{\theta}}-\ve\theta)^T)$, and in (\ref{eq:cs2}) we select $\ve b = \ve i_\text{MP}^{-1}(\ve\theta)\ve a$. As a consequence of assumption iii), $\ve i_\text{MP}^{-1}(\ve\theta)$ is positive-definite, $\ve a\ve b^T\geq 0$, and from (\ref{eq:cs2}) we have that $\ve a^T(\mathbf{C}_\text{MP}-\ve i_\text{MP}^{-1}(\ve\theta))\ve a\geq 0, \forall \ve a$. Now, the equality in (\ref{eq:cs0}) holds if and only if
$\ve a^T (\hat{\ve{\theta}}-\ve\theta) = c\,\ve a^T \ve i_\text{MP}^{-1}(\ve\theta) \frac{\partial\log p_\text{MP}(\ve z;\ve{\theta})}{\partial\ve{\theta}}
$. As this is satisfied $\forall \ve a$, we finally obtain (\ref{eq:facMP}) given that the constant $c$ is proved to be $1$. 

The second part of the proof is as follows:
by consistency of $\hat{\ve{\theta}}$, $(\ref{eq:facMP})$ is also satisfied with $\hat{\ve\theta}$ instead of $\ve\theta$ when $L\rightarrow\infty$. 
Then, using a first-order Taylor expansion of $\ve i(\ve{\theta})$ around $\hat{\ve{\theta}}$ and discarding the second order terms as $L\rightarrow\infty$, we have  
$
\textstyle\frac{\partial\log p_\text{MP}(\ve z;\ve{\theta})}{\partial \ve\theta} = L \ve i_\text{MP}(\hat{\ve\theta})(\hat{\ve\theta}-\ve\theta ).
$ Integrating this equation with respect to $\ve \theta$:
\begin{align}
\textstyle\log p_\text{MP}(\ve z;\ve\theta ) &=\textstyle -\frac{L}{2} (\hat{\ve\theta}-\ve\theta )^T \ve i_\text{MP}(\hat{\ve\theta})(\hat{\ve\theta}-\ve\theta )+ c(\hat{\ve\theta}),
\label{eq:logp}
\end{align}
where the integration constant must be $c(\hat{\ve\theta}) = \log p_\text{MP}(\ve z;\hat{\ve\theta})$ given that (\ref{eq:logp}) is satisfied asymptotically by the consistency of $\hat{\ve\theta}$ when $L\rightarrow\infty$. Therefore, 
$ \textstyle p_\text{MP}(\ve z;\ve\theta) = p_\text{MP}(\ve z;\hat{\ve\theta}) e^{-L\frac{1}{2} \left(\hat{\ve\theta}-\ve\theta \right)^T \ve i_\text{MP}(\hat{\ve\theta})\left(\hat{\ve\theta}-\ve\theta \right)}.
$ Using the previous equation in the expression of $T_\text{L-MP}$ and replacing $\ve\theta$ by $\ve \theta^0$, we obtain
$T_\text{L-MP}(\ve z)=\frac{ p_\text{MP}(\ve z;\hat{\ve\theta})}{ p_\text{MP}\left(\ve z;\ve\theta^0\right)} = e^{L\frac{1}{2} \left(\hat{\ve\theta} - \ve{\theta^0}\right)^T \ve i_\text{MP}(\hat{\ve\theta})\left(\hat{\ve\theta}-\ve{\theta^0}\right) },
$ or 
\begin{align*}
\textstyle 2\log T_\text{L-MP}(\ve z)&= L (\hat{\ve\theta} - \ve{\theta^0})^T \ve i_\text{MP}(\hat{\ve\theta})(\hat{\ve\theta}-\ve{\theta^0}) 
\end{align*}
Using again the CMT and the continuity of the second-order partial derivatives, the following is satisfied when $L\rightarrow\infty$:
$\ve i_\text{MP}(\hat{\ve\theta})(\hat{\ve\theta}-\ve{\theta}^0) =
\ve i_\text{MP}(\ve\theta^i)(\hat{\ve\theta}-\ve{\theta}^0) \ \text{under } \Hip_i, i=0,1.
$
Finally, 
\begin{align}
\textstyle 2\log T_\text{L-MP}(\ve z)&=\textstyle L (\hat{\ve\theta}-\ve\theta^0)^T \ve i_\text{MP}(\ve \theta^i)(\hat{\ve\theta}-\ve\theta^0)\\
&= \textstyle\|\sqrt{L}\ve i_\text{MP}^{\frac{1}{2}}(\ve \theta^i)(\hat{\ve\theta}-\ve\theta^0)\|^2\\ &\overset{a}{\sim} f_P(\ve{\mu_{\text{MP},i},\ve{\Sigma_{\text{MP},i}}}) \text{ under } \Hip_i.
\label{eq:TLdist}
\end{align}
where the parameters $\ve \mu_{\text{MP},i}$ and $\ve\Sigma_{\text{MP},i}$ of the asymptotic distribution $f_P$ are the mean and the covariance matrix of the multivariate Gaussian vector inside the square norm. They are obtained using (\ref{eq:lmle}) and are presented in the lemma. 

\fi
 
\bibliographystyle{unsrt}
\bibliography{IEEEabrv,../../Refs/refs}

\begin{thebibliography}{10}

\bibitem{shaikh2016energy}
Faisal~Karim Shaikh and Sherali Zeadally.
\newblock Energy harvesting in wireless sensor networks: A comprehensive
  review.
\newblock {\em Renewable and Sustainable Energy Reviews}, 55:1041--1054, 2016.

\bibitem{rashid2016applications}
Bushra Rashid and Mubashir~Husain Rehmani.
\newblock Applications of wireless sensor networks for urban areas: A survey.
\newblock {\em Journal of Network and Computer Applications}, 60:192--219,
  2016.

\bibitem{li2020distributed}
F.~{Li}, M.~{Valero}, Y.~{Cheng}, T.~{Bai}, and W.~{Song}.
\newblock Distributed sensor networks based shallow subsurface imaging and
  infrastructure monitoring.
\newblock {\em IEEE Transactions on Signal and Information Processing over
  Networks}, 6:241--250, 2020.

\bibitem{gupta2020collaborative}
V.~{Gupta} and S.~{De}.
\newblock Collaborative multi-sensing in energy harvesting wireless sensor
  networks.
\newblock {\em IEEE Transactions on Signal and Information Processing over
  Networks}, 6:426--441, 2020.

\bibitem{gubbi2013internet}
J.~Gubbi, R.~Buyya, S.~Marusic, and M.~Palaniswami.
\newblock {Internet of Things (IoT): A vision, architectural elements, and
  future directions}.
\newblock {\em Future generation computer systems}, 29(7):1645--1660, 2013.

\bibitem{al2015internet}
Ala Al-Fuqaha, Mohsen Guizani, Mehdi Mohammadi, Mohammed Aledhari, and Moussa
  Ayyash.
\newblock Internet of things: A survey on enabling technologies, protocols, and
  applications.
\newblock {\em IEEE Communications Surveys \& Tutorials}, 17(4):2347--2376,
  2015.

\bibitem{chepuri2016sparse}
Sundeep~Prabhakar Chepuri and Geert Leus.
\newblock Sparse sensing for distributed detection.
\newblock {\em IEEE Transactions on Signal Processing}, 64(6):1446--1460, 2016.

\bibitem{ciuonzo2017distributed}
D.~Ciuonzo and P.~S. Rossi.
\newblock Distributed detection of a non-cooperative target via generalized
  locally-optimum approaches.
\newblock {\em Information Fusion}, 36:261--274, 2017.

\bibitem{aldalahmeh2019fusion}
Sami~Ahmed Aldalahmeh, Saleh~O Al-Jazzar, Des McLernon, Syed Ali~Raza Zaidi,
  and Mounir Ghogho.
\newblock Fusion rules for distributed detection in clustered wireless sensor
  networks with imperfect channels.
\newblock {\em IEEE Transactions on Signal and Information Processing over
  Networks}, 5(3):585--597, 2019.

\bibitem{ciuonzo2014decision}
D.~Ciuonzo and P.S. Rossi.
\newblock Decision fusion with unknown sensor detection probability.
\newblock {\em IEEE Signal Processing Letters}, 21(2):208--212, 2014.

\bibitem{hamed2012reliable}
S~Hamed Hamed and Ali Peiravi.
\newblock Reliable distributed detection in multi-hop clustered wireless sensor
  networks.
\newblock {\em IET Signal Processing}, 6(8):743--750, 2012.

\bibitem{kar2011distributed}
Soummya Kar, Ravi Tandon, H~Vincent Poor, and Shuguang Cui.
\newblock Distributed detection in noisy sensor networks.
\newblock In {\em Information Theory Proceedings (ISIT), 2011 IEEE
  International Symposium on}, pages 2856--2860. IEEE, 2011.

\bibitem{li2018fully}
Shang Li and Xiaodong Wang.
\newblock Fully distributed sequential hypothesis testing: Algorithms and
  asymptotic analyses.
\newblock {\em IEEE Transactions on Information Theory}, 64(4):2742--2758,
  2018.

\bibitem{Ciuonzo_Rossi_Willett_2017}
Domenico Ciuonzo, Pierluigi~Salvo Rossi, and Peter Willett.
\newblock Generalized rao test for decentralized detection of an uncooperative
  target.
\newblock {\em IEEE Signal Processing Letters}, 24(5):678--682, May 2017.

\bibitem{ghasemi2007asymptotic}
Amir Ghasemi and Elvino~S Sousa.
\newblock Asymptotic performance of collaborative spectrum sensing under
  correlated log-normal shadowing.
\newblock {\em Communications Letters, IEEE}, 11(1):34--36, 2007.

\bibitem{sayed2014adaptation}
Ali~H Sayed et~al.
\newblock Adaptation, learning, and optimization over networks.
\newblock {\em Foundations and Trends{\textregistered} in Machine Learning},
  7(4-5):311--801, 2014.

\bibitem{al2018node}
S.~Al-Sayed, J.~Plata-Chaves, M.~Muma, M.~Moonen, and A.M. Zoubir.
\newblock Node-specific diffusion lms-based distributed detection over adaptive
  networks.
\newblock {\em IEEE Trans. on Sig. Proc.}, 66(3):682--697, 2018.

\bibitem{sayed2013diffusion}
Ali~H Sayed, Sheng-Yuan Tu, Jianshu Chen, Xiaochuan Zhao, and Zaid~J Towfic.
\newblock Diffusion strategies for adaptation and learning over networks: an
  examination of distributed strategies and network behavior.
\newblock {\em IEEE Signal Processing Magazine}, 30(3):155--171, 2013.

\bibitem{Sahu_Kar_2017}
Anit~Kumar Sahu and Soummya Kar.
\newblock Recursive distributed detection for composite hypothesis testing:
  Nonlinear observation models in additive gaussian noise.
\newblock {\em IEEE Transactions on Information Theory}, 63(8):4797--4828, Aug
  2017.

\bibitem{cattivelli2011distributed}
Federico~S Cattivelli and Ali~H Sayed.
\newblock Distributed detection over adaptive networks using diffusion
  adaptation.
\newblock {\em IEEE Transactions on Signal Processing}, 59(5):1917--1932, 2011.

\bibitem{drakopoulos1991optimum}
Elias Drakopoulos and C-C Lee.
\newblock Optimum multisensor fusion of correlated local decisions.
\newblock {\em IEEE Transactions on Aerospace and Electronic Systems},
  27(4):593--606, 1991.

\bibitem{villas2014spatial}
Leandro~A Villas, Azzedine Boukerche, Horacio~ABF De~Oliveira, Regina~B
  De~Araujo, and Antonio~AF Loureiro.
\newblock A spatial correlation aware algorithm to perform efficient data
  collection in wireless sensor networks.
\newblock {\em Ad Hoc Networks}, 12:69--85, 2014.

\bibitem{Wang_Reiss_Cavallaro_2016}
L.~Wang, J.~D. Reiss, and A.~Cavallaro.
\newblock Over-determined source separation and localization using distributed
  microphones.
\newblock {\em IEEE/ACM Transactions on Audio, Speech, and Language
  Processing}, 24(9):1573--1588, Sep 2016.

\bibitem{chen2019layered}
Siguang Chen, Shujun Zhang, Xiaoyao Zheng, and Xiukai Ruan.
\newblock Layered adaptive compression design for efficient data collection in
  industrial wireless sensor networks.
\newblock {\em Journal of Network and Computer Applications}, 129:37--45, 2019.

\bibitem{lunden2015spectrum}
Jarmo Lunden, Visa Koivunen, and H~Vincent Poor.
\newblock Spectrum exploration and exploitation for cognitive radio: Recent
  advances.
\newblock {\em IEEE signal processing magazine}, 32(3):123--140, 2015.

\bibitem{Liang_Zeng_Peh_Hoang_2008}
Y.~C. Liang, Y.~Zeng, E.~C.~Y. Peh, and A.~T. Hoang.
\newblock Sensing-throughput tradeoff for cognitive radio networks.
\newblock {\em IEEE Transactions on Wireless Communications}, 7(4):1326--1337,
  Apr 2008.

\bibitem{Tong_2007}
A.~Anandkumar, L.~Tong, and A.~Swami.
\newblock {Detection of Gauss-Markov Random Fields With Nearest-Neighbor
  Dependency}.
\newblock {\em {IEEE} Trans. Inf. Theory}, 55(2):816--827, Feb 2009.

\bibitem{anandkumar2009scalable}
Animashree Anandkumar.
\newblock {\em Scalable algorithms for distributed statistical inference}.
\newblock PhD thesis, Cornell University, 2009.

\bibitem{Lin_Chen_2015}
Ying Lin and Hao Chen.
\newblock Distributed detection performance under dependent observations and
  nonideal channels.
\newblock {\em IEEE Sensors Journal}, 15(2):715--722, Feb 2015.

\bibitem{nelsen2007introduction}
Roger~B Nelsen.
\newblock {\em An introduction to copulas}.
\newblock Springer Science \& Business Media, 2007.

\bibitem{he2015fusing}
Hao He and Pramod~K Varshney.
\newblock Fusing censored dependent data for distributed detection.
\newblock {\em IEEE Transactions on Signal Processing}, 63(16):4385--4395,
  2015.

\bibitem{iyengar20178}
Satish~G Iyengar, Hao He, Arun Subramanian, Ruixin Niu, Pramod~K Varshney, and
  Thyagaraju Damarla.
\newblock 8 distributed detection and data fusion with heterogeneous sensors.
\newblock {\em Multisensor Data Fusion: From Algorithms and Architectural
  Design to Applications}, page 127, 2017.

\bibitem{zhang_multi-antenna_2010}
Rui Zhang, Teng~Joon Lim, Ying-Chang Liang, and Yonghong Zeng.
\newblock Multi-antenna based spectrum sensing for cognitive radios: {A} {GLRT}
  approach.
\newblock {\em IEEE Transactions on Communications}, 58(1):84--88, January
  2010.
\newblock Conference Name: IEEE Transactions on Communications.

\bibitem{taherpour_multiple_2010}
Abbas Taherpour, Masoumeh Nasiri-Kenari, and Saeed Gazor.
\newblock Multiple antenna spectrum sensing in cognitive radios.
\newblock {\em IEEE Transactions on Wireless Communications}, 9(2):814--823,
  February 2010.
\newblock Conference Name: IEEE Transactions on Wireless Communications.

\bibitem{penna_decentralized_2015}
Federico Penna and Slawomir Stanczak.
\newblock Decentralized {Eigenvalue} {Algorithms} for {Distributed} {Signal}
  {Detection} in {Wireless} {Networks}.
\newblock {\em IEEE Transactions on Signal Processing}, 63(2):427--440, January
  2015.

\bibitem{javadi2016detection}
S~Hamed Javadi.
\newblock Detection over sensor networks: a tutorial.
\newblock {\em IEEE Aerospace and Electronic Systems Magazine}, 31(3):2--18,
  2016.

\bibitem{Kay_SSP}
S.~Kay.
\newblock {\em Fundamentals of Statistical Signal Processing, Volume II:
  Detection Theory}.
\newblock Prentice-Hall, 1st ed. edition, 1998.

\bibitem{Levy_Det}
Bernard~C Levy.
\newblock {\em {Principles of signal detection and parameter estimation}}.
\newblock Springer, 2008.

\bibitem{Kay_SSP_ET}
Steven~M Kay.
\newblock {\em Fundamentals of statistical signal processing, Volume I:
  Estimation theory}.
\newblock Prentice-Hall, Inc., 1993.

\bibitem{xiao2004fast}
Lin Xiao and Stephen Boyd.
\newblock Fast linear iterations for distributed averaging.
\newblock {\em Systems \& Control Letters}, 53(1):65--78, 2004.

\bibitem{lugannani1980saddle}
Robert Lugannani and Stephen Rice.
\newblock Saddle point approximation for the distribution of the sum of
  independent random variables.
\newblock {\em Advances in applied probability}, pages 475--490, 1980.

\bibitem{Gudmundson1991}
M.~Gudmundson.
\newblock Correlation model for shadow fading in mobile radio systems.
\newblock {\em Electronics Letters}, 27(23):2145--2146, Nov 1991.

\bibitem{zhou2011modeling}
H.Y. Zhou, D.Y. Luo, Y.~Gao, and D.C. Zuo.
\newblock Modeling of node energy consumption for wireless sensor networks.
\newblock {\em Wireless Sensor Network}, 3(1):18, 2011.

\bibitem{kshemkalyani2011distributed}
Ajay~D Kshemkalyani and Mukesh Singhal.
\newblock {\em Distributed computing: principles, algorithms, and systems}.
\newblock Cambridge University Press, 2011.

\bibitem{sundaram_finite-time_2007}
Shreyas Sundaram and Christoforos~N. Hadjicostis.
\newblock Finite-{Time} {Distributed} {Consensus} in {Graphs} with
  {Time}-{Invariant} {Topologies}.
\newblock In {\em 2007 {American} {Control} {Conference}}, pages 711--716, July
  2007.
\newblock ISSN: 2378-5861.

\bibitem{blatt2004distributed}
Doron Blatt and Alfred Hero.
\newblock Distributed maximum likelihood estimation for sensor networks.
\newblock In {\em 2004 IEEE International Conference on Acoustics, Speech, and
  Signal Processing}, volume~3, pages iii--929. IEEE, 2004.

\bibitem{wiesel2011distributed}
Ami Wiesel and Alfred~O Hero.
\newblock Distributed covariance estimation in gaussian graphical models.
\newblock {\em IEEE Transactions on Signal Processing}, 60(1):211--220, 2011.

\bibitem{2015PoorSpectrumSensing}
J.~Lunden, V.~Koivunen, and H.V. Poor.
\newblock Spectrum exploration and exploitation for cognitive radio: Recent
  advances.
\newblock {\em Signal Processing Magazine, IEEE}, 32(3):123--140, May 2015.

\bibitem{zeng2008blindly}
Yonghong Zeng, Ying-Chang Liang, and Rui Zhang.
\newblock Blindly combined energy detection for spectrum sensing in cognitive
  radio.
\newblock {\em IEEE Signal Processing Letters}, 15:649--652, 2008.

\bibitem{bianchi_performance_2011}
P.~Bianchi, M.~Debbah, M.~Maida, and J.~Najim.
\newblock Performance of {Statistical} {Tests} for {Single}-{Source}
  {Detection} {Using} {Random} {Matrix} {Theory}.
\newblock {\em IEEE Transactions on Information Theory}, 57(4):2400--2419,
  April 2011.
\newblock Conference Name: IEEE Transactions on Information Theory.

\bibitem{apostol1974mathematical}
Tom~M. Apostol.
\newblock {\em Mathematical analysis}, volume~2.
\newblock Addison-Wesley Reading, MA, 1974.

\bibitem{Vaart_2000}
A.~W. van~der Vaart.
\newblock {\em Asymptotic Statistics}.
\newblock Cambridge University Press, Jun 2000.

\end{thebibliography}
 \end{document}